\newcounter{jvcc}

\newcounter{jjmc}

\documentclass[%
 reprint,
superscriptaddress,
 amsmath,amssymb, aps,
]{revtex4-2}

\newcommand{\tr}{\mathrm{tr}} \newcommand{\str}{\mathrm{str}}
\newcommand{\STr}{\mathrm{STr}}

\usepackage{graphicx}
\usepackage{dcolumn}
\usepackage{bm}
\usepackage{dsfont}
\usepackage[export]{adjustbox}
\usepackage{comment}
\usepackage[normalem]{ulem}
\usepackage{xcolor}

\usepackage{graphicx,color,hyperref}

\begin{document}

\preprint{APS/123-QED}

\title{Quantum Chaos on Edge}

\author{Alexander Altland}
 \affiliation{Institut für theoretische Physik, Zülpicher Str. 77, 50937 Köln, Germany}

 \author{Kun Woo Kim}%
 \affiliation{Department of Physics, Chung-Ang University, 06974 Seoul, Republic of Korea}

 \author{Tobias Micklitz}%
 \affiliation{Centro Brasileiro de Pesquisas Físicas, Rua Xavier Sigaud 150, 22290-180, Rio de Janeiro, Brazil}

 \author{Maedeh Rezaei}%
 \affiliation{Institut für theoretische Physik, Zülpicher Str. 77, 50937 Köln, Germany}

 \author{Julian Sonner}%
 \affiliation{Department of Theoretical Physics, University of Geneva, 24 quai
 Ernest-Ansermet,  1211 Gen\`eve 4, Suisse}%

 \author{Jacobus J. M. Verbaarschot}%
 \affiliation{ Center for Nuclear Theory and Department of Physics and Astronomy, Stony Brook University, Stony Brook, New York 11794, USA}

\date{\today}

\begin{abstract}
In recent years, the physics of many-body quantum chaotic systems close to their
ground states has come under intensified scrutiny. Such studies are motivated by
the emergence of model systems exhibiting chaotic fluctuations throughout the
entire spectrum (the Sachdev-Ye-Kitaev (SYK) model being a renowned
representative) as well as by the physics of holographic principles, which
likewise unfold close to ground states. Interpreting the edge of the spectrum as
a quantum critical point, here we combine a wide range of analytical and
numerical methods to the identification and comprehensive description of two
different universality classes: the near edge physics of  ``sparse'' and the
near edge of ``dense'' chaotic systems. The distinction lies in the ratio
between the number of a system's random parameters and its Hilbert space
dimension,  which is exponentially small or algebraically small  in the sparse
and dense case, respectively. Notable representatives of the two classes are
generic chaotic many-body models (sparse) and single particle systems, invariant
random matrix ensembles, or chaotic gravitational systems (dense). While the two
families share identical spectral correlations at energy scales comparable to
the level spacing, the density of states and its fluctuations near the edge are
different. Considering  the SYK model as a representative of the sparse class,
we apply a combination of field theory and exact diagonalization to a detailed
discussion of its edge spectrum. Conversely, Jackiw-Teitelboim gravity is our
reference model for the dense class, where an analysis of the gravitational path
integral and random matrix theory reveal universal differences to the sparse
class, whose implications for the construction of holographic principles we
discuss.

\end{abstract}

\maketitle


\section{Introduction}

\emph{Random matrix universality} represents one of the most powerful
universality principles in quantum physics: Correlations between close-by energy
levels of quantum systems which are chaotic (and ergodic) in their long time
limit are statistically equivalent to those of Gaussian distributed random
matrices. Depending on the system class under consideration, a spectrum of
different methods is available to establish this correspondence, including
periodic orbit theory~\cite{Haake}, methods of quantum field
theory~\cite{Efetbook}, matrix theory~\cite{MehtaRMT}, and most recently the
analysis of gravitational path integrals~\cite{Saad2019}. 

Spectral analysis  often focuses on quantum states deep inside the spectrum,
where the average spectral density is structureless, or even approximately
uniform on small energy scales.  In which  ways do correlations between neighboring levels
change upon approaching the edge of a spectrum?  Asking this question is
motivated by recent developments in quantum many body physics and in gravity. In
the former context, one is often interested in physics close to a many-body
ground state. While Hamiltonian dynamics at low excitation energies generically
is integrable, some system classes remain chaotic all the way down\footnote{Without entering the subtle question of defining criteria for `chaos', we consider \emph{ergodicity} of quantum states (in the sense of Ref.~\cite{monteiroQuantumErgodicityManybody2020}) as a necessary condition. Examples of phases ruled out on this basis include (many body) localized, or glassy phases.}. A celebrated
example is the SYK model, i.e. a maximally random  pair interaction model of
(Majorana) fermions \cite{kitaevSimpleModelQuantum2015}. Another one is
two-dimensional Jackiv-Teitelboim (JT) gravity (see Ref.~\cite{mertensJT2023}
for review). JT gravity satisfies a holographic principle, in which it becomes
the bulk dual of a one-dimensional quantum chaotic boundary theory close to its
ground state. The description of this correspondence requires a fine-grained
statistical resolution of this spectral edge. 

Random matrix theory itself has no difficulties with the description of spectra
near the edge, including extreme value statistics (the probability distribution
of the lowest eigenvalue of the matrix ensemble), spectral correlations, and the
detailed profile of the average near-edge spectral
density \cite{MehtaRMT,TracyWidomAiry1993}.  However, for `real' systems, the
universality principle  is up for renegotiation: are systems which display
quantum chaos all the way down to their edge (in a sense to be made precise
below) universally described by the random matrix paradigm? If not, can we
identify distinct universality classes? Which of the above theoretical
frameworks remain applicable in the immediate vicinity of the edge, i.e. how can
we quantitatively describe observables such as spectral densities and their
correlations beyond random matrix theory?

In this paper, we approach these questions from different perspectives, the most general one being 
an interpretation of the edge as a quantum critical point of a
symmetry breaking quantum phase transition. The control parameter of this
transition is energy, $\epsilon$, and its order parameter is the  ensemble
average $\langle \rho(\epsilon)\rangle$ of the spectral density
\begin{align}
    \label{eq:RhoResolvent}
    \rho(\epsilon )= -\frac{1}{2\pi i} \tr(G(\epsilon^+)-G(\epsilon^-)), 
\end{align}        
where $G(z)\equiv (z-H)^{-1}$ is the resolvent,  $H$ the system Hamiltonian,
$\epsilon^\pm = \epsilon \pm i0$,   and $\langle \dots \rangle$ an average over
microscopically different realizations of $H$. A spectral edge at
$\epsilon_\textrm{e}\equiv 0$ is characterized by a profile  $\langle \rho(\epsilon)\rangle\sim
\Theta(\epsilon)\epsilon^\alpha$, $0<\alpha<1$, resembling that of, for example, the
magnetization at a ferromagnetic transition. (The transition is  symmetry
breaking in that for each realization $H$ the
block diagonal operator $\hat{G}\equiv (\epsilon + i 0 \tau_3 - H)^{-1}$ is
invariant under continuous rotations in the two-dimensional `causal'
representation space of the Pauli matrix $\tau_3$. However, the ensemble
average $\rho\to \langle \rho \rangle$ breaks the symmetry at
energies $\epsilon>0$ via the onset of a finite density of states Eq.~\eqref{eq:RhoResolvent}, i.e. the degeneracy between $G^+ $ and $ G^- $ is lifted and the symmetry broken.)

The understanding of the edge as a phase transition point unlocks an arsenal of
concepts and methods of statistical mechanics in the analysis of the edge
problem.  Specifically, we know that upon approaching a quantum critical point
in terms of the control parameter $\epsilon$, the theory must develop `gapless
fluctuations', i.e. fluctuations that become unbound in the limit $\epsilon \to
0$. Below, we will reason that there are at least two universality classes of
many-body quantum chaotic models distinguished by different fluctuation
behavior, and accordingly different edge phenomenology.

\begin{figure}[htbp]
    \centering
    \includegraphics[width=0.8\columnwidth]{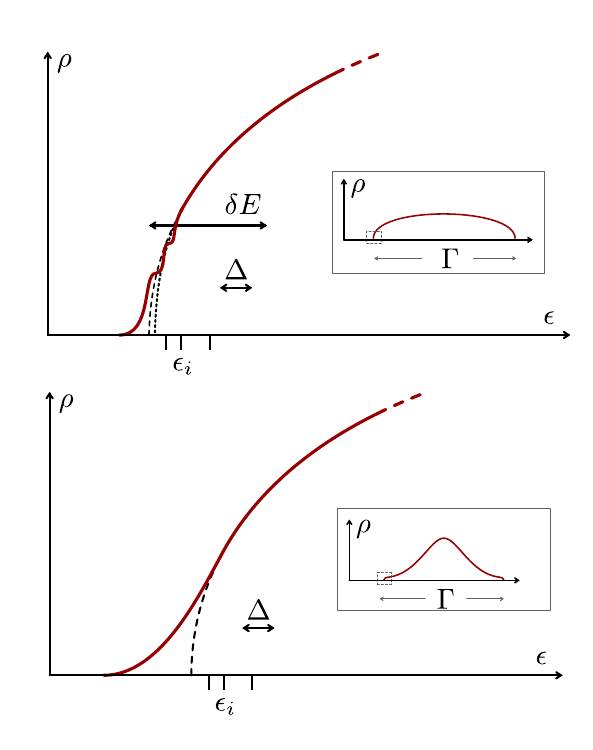}
    \caption{Upper panel: near edge spectral density of dense systems. A small
    tail and oscillations periodic in $\sim D^{-1}$ well away from the edge
    superimposed on the $\sim
    \epsilon^{1 /2}$ (long dashed) mean field profile indicate an almost
    crystalline ordering of levels close to the edge. Inside the edge, there
    exists a weak universal power law correction $\sim \epsilon^{-5 /2}$ (short
    dashed), which may serve as a perturbative indicator of the dense
    universality class in physical models such as JT gravity. Bottom: the near
    edge spectral density of sparse models is softer, structures at level
    spacing scales are washed out. The spectral density at large scales (inset),
    too, is broader distributed than it is in the dense case.
    Here, $\Gamma$ and $\Delta$ are the band width and mean level spacing, respectively, 
    and $\delta E$ is the energy range over which the scaling form of the density of states holds (see also discussion in main text).}
    \label{fig:EdgeComparison}
\end{figure}

\paragraph*{Dense systems:} We call systems \emph{dense} if their  number of
independent random system parameters is roughly of the same order as the Hilbert
space dimension, $D$. Representatives of this class include synthetic models such as
random matrix ensembles or random graphs,  and, as we will discuss, JT gravity.
The edge of dense models turns out to be hard in that corrections to the above
mean field power law are small in $D^{-1}$~\footnote{Confusingly, the random
matrix community calls such edges 'soft'. However, for us they are `hard' in
comparison to the much softer edges of other model systems.}. These
modifications include an exponentially small tail of spectral density
$\rho(\epsilon)$ leaking beyond the edge for $\epsilon<0$, and a  power law
correction proportional to $(D \epsilon)^{-5 /2}$ to the leading $\sim
\epsilon^{1 /2}$ inside the region of spectral support, $\epsilon>0$.
Superimposed on this  envelope, there are oscillations periodic in the  level
spacing indicating almost crystalline order in the level position near the hard
edge (cf. Fig.~\ref{fig:EdgeComparison}). 

While these are known signatures of invariant random matrix
ensembles~\cite{TracyWidomAiry1993}, our discussion here is geared towards
systems which are defined differently, and to which the toolbox of random matrix
methodology does not immediately apply. Instead, we will demonstrate that dense
systems define a broad universality class whose near edge physics is
quantitatively described by  an effective (low-dimensional) matrix theory  known as the Kontsevich
model~\cite{kontsevich1992}. From this model, the universal features of the
dense spectrum can be conveniently described, as we will demonstrate. 

\paragraph*{Sparse systems:} By our definition, sparse systems contain at most
$\mathcal{O}(\ln(D))$ independent random parameters. This is the situation
generically realized in randomly interacting many-body systems. In these
systems, we have $N$ degrees of freedom, think of spins or qubits, defining a
Hilbert space of dimension $D\sim 2^N$ . Assuming few-body interactions, the
associated Hamiltonians contain random parameters whose number grows
polynomially in $N\sim \ln D$. In a first quantized representation, these
Hamiltonians are  matrices which are sparse in that they contain only
$\mathcal{O}(D\log D)$.  (not $\mathcal{O}(D^2)$) of
non-vanishing matrix elements, and massively correlated, in that there are only
$\mathcal{O}(\ln D)$ statistically independent parameters.    

Referring for a detailed discussion to section ~\ref{sec:Sparse}, these
differences affect the structure of spectra all the way from scales comparable
to the band width down to the correlations of individual levels. In particular,
the edge turns out to be much softer than in the dense case. The tails leaking
beyond the $ D\to \infty$ mean field edge, now are of extent $\sim \ln(D)^{-1}$,
parametrically exceeding the level spacing. There are no superimposed
oscillatory structures and no power law corrections to the mean field limit
inside the edge. We note that the formation of a comparatively soft edge does
not contradict the interpretation in terms of a quantum critical point: writing
the spectral density as $\langle \rho \rangle= \bar \rho(1+\dots)$ all terms
represented by  ellipses vanish in the `thermodynamic limit', $D \to \infty$,
and the spectral density of both, sparse and dense systems retracts to the  mean
field order parameter $\bar \rho\equiv  \epsilon^\alpha \Theta(\epsilon) $
heralding the phase transition. (In the qualitative representation of figure
\ref{fig:EdgeComparison}, this implies that the deviations from the long dashed
line representing \(\bar \rho\) become  small in the same sense.)  For the same reason,  our two classes sparse and dense do
not represent distinct \emph{phases}; They are universality classes associated
to signatures visible in the regime of large but finite $D$.

In section \ref{sec:Sparse}, we will discuss how the differences to the dense
case originate in the presence of two channels of correlations in sparse
systems. The first of these, can be traced back to statistical dependence: turning only few `knobs' in the Hamiltonian affects an exponentially
large number of matrix elements, and in this way generates large scale `collective' fluctuations in the spectrum. The second describes the microscale
correlations of nearby levels, as in dense systems. Empirically, we know that
these two types of fluctuations operate largely independently of each other.
In models affording a high degree of analytical solvability,
collective~\cite{Garcia-Garcia2016,garcia-garciaAnalyticalSpectralDensity2017}
and microscale~\cite{altlandQuantumErgodicitySYK2018} fluctuations can be
treated by tailored methods assuming the form of, e.g., summations over
different perturbative diagram classes. Considering the SYK model as a case
study, we will here introduce a comprehensive framework in which the
fluctuations of a sparse system are treated in a unified fashion. Zooming in to
the edge, we will show how the theory assumes the form of a matrix theory
different from the Kontsevich model. 

We reason that the emergence of two different effective theories reflects the
existence of at least two universality classes for the edge transition, dense
and sparse. The differences between these two classes become of importance in
cases where physical phenomena close to quantum ground states are addressed. As
a case study, we will consider the two-dimensional holographic correspondence,
which motivated the introduction of the SYK model in the first
place~\cite{kitaevSimpleModelQuantum2015}. In this context, JT gravity and the
SYK model feature representatives of the dense and sparse class, respectively.
(The intuition behind the gravity setting in the dense class is that the
gravitational path integral describes fluctuating geometries. Thinking of the
latter in terms of discrete tessellations of space-time, 
fluctuating matrix structures
containing about as many random parameters as degrees of freedom emerge --- a
dense scenario.) The finding that the two models fall into different
universality classes may be somewhat sobering news for the construction of
holographic correspondences relying on principles of chaos: In the regimes of
interest, so-called double scaled energy windows close to the band edge, the
differences in the structure of bulk (JT) and boundary (SYK) spectra show in the
average spectral density as well as in its statistical correlations.
Specifically, we will demonstrate that the collective fluctuations affect the spectral
form factor (Eq.~\eqref{eq:SpectralFormFactorDef}), including at time scales
$\tau \gtrsim 1$, where correlations in the spectrum at level spacing scales are
probed. (The statistical independence of collective and micro-scale fluctuations
is not in conflict with the former influencing the latter down to the smallest
scales.) A fully developed holographic correspondence would not tolerate
such differences, and hence require the model sitting in the dense class at the
boundary. However, this criterion is not realizable in terms of `natural'
theories defined in terms of a few-body interaction. Instead, one needs to resort
to more artificial constructions such as the so-called double scaled SYK
model~\cite{Cotler2017}, where $q$-body interactions with $q$ scaling as a power
of $N$, define a dense setting. In section \ref{sec:Numerical}, we will
demonstrate how the structure of the edge interpolates between the sparse and
the dense profiles as we increase the large-$N$ scaling of $q$ in an SYK Hamiltonian.    

The remainder of the paper is organized as follows: We start in section
\ref{sec:Scene} with a more precise definition of the spectral edge, and quick
introduction of statistical diagnostics that will be used throughout. In
sections \ref{sec:Dense} and \ref{sec:Sparse} we will discuss the near edge
physics of dense and sparse systems, respectively. We will compare our
predictions to exact diagonalization in section \ref{sec:Numerical}, and in
section \ref{sec:Gravity} discuss two-dimensional holography as an application.
Several more technical discussions, such as the first principle derivation of
the effective edge theories for random matrix Hamiltonian (dense) or the SYK
Hamiltonian (sparse) are relegated to Appendices.

\section{Setting the scene}
\label{sec:Scene}

We start our discussion with a  sharpened definition of the terminology
`spectral edge'.  Consider an ensemble of chaotic quantum systems defined in a
$D$-dimensional Hilbert space  whose spectral density vanishes on average as
\(\langle \rho(\epsilon) \rangle \approx \bar \rho\sim  \epsilon^{\alpha}
\,\Theta(\epsilon)\), with \(0<\alpha<1\).  Here, the `mean field' approximation
$\bar \rho$  is defined to exclude contributions to $\langle \rho \rangle$
vanishing in the limit $D\to \infty$. (Throughout this paper, spectral densities
averaged in different ways will appear frequently. Unless stated otherwise, we
use the notation $\bar \rho$ for the leading order contribution to the spectral
density and $\langle \rho \rangle$ for the exact ensemble average. Occasionally,
we will need to indicate the type of averaging, such as $\langle \dots
\rangle_\textrm{coll}$ for the average over `collective fluctuations in sparse
systems'. If no confusion is possible, we omit the energy argument, $\bar
\rho(\epsilon)=\bar \rho $, etc.) We implicitly define the region of energies
over which the above scaling form holds as $\delta E \ll \Gamma      $, where
$\Gamma$ is the band width of the system. Deep inside the spectrum, individual
energy levels are separated by a spacing $\Delta = \langle \rho \rangle
^{-1}\sim \Gamma / D$. Upon approaching the edge, the level spacing increases as
$\Delta(\epsilon)\sim \Delta_0(\Delta_0 /\epsilon )^\alpha$, where  
$\Delta_0$ is the `largest' average 
level spacing in the system, i.e. the
spacing between the two outermost levels, and implicitly defined by the
condition $\int_0^{\Delta_0} d \epsilon \Delta(\epsilon)^{-1}=1 $.      

The systems considered here are chaotic down to the edge in the sense that for
 all energies satisfying the condition $\epsilon \gg \Delta(\epsilon)$ spectral \emph{correlations} are
governed by the universal correlation functions of random matrix theory.
For Hamiltonians not possessing antilinear symmetries besides
hermiticity -- the symmetry class considered in this paper --, this 
implies that the connected two-point correlation function
\begin{align}
    \label{eq:SpectralTwoPointDef}
    C(\epsilon_1, \epsilon_2
)\equiv \Delta^2(\epsilon)\left\langle
\rho\left(\epsilon_1 \right)\rho\left(\epsilon_2 \right)
\right\rangle_\textrm{c}, 
\end{align}
is given by~\cite{MehtaRMT}
\begin{align}
    \label{eq:R2SineKernel}
    C(\epsilon_1,\epsilon_2 )=- \left \langle \left( \frac{\sin s}{s} \right)^2 \right \rangle_\textrm{coll},\qquad s \equiv \pi \omega \langle \rho(\epsilon) \rangle,
\end{align} 
where $\epsilon=(\epsilon_1 + \epsilon_2 )/ 2$ and $\omega=\epsilon_1 - \epsilon_2 $.  
The physical meaning of the  oscillatory  
    correlation function is level rigidity, i.e. an almost uniform spacing of levels
    with period $\Delta(\epsilon)$. For later reference, we note that this result can be equivalently represented as 
    \begin{align}
        \label{eq:CorrelationFunctionSineKernel}
        C(\epsilon_1 ,\epsilon_2 )=- K^2_\textrm{sin}\left( \rho \epsilon_1 ,\rho\epsilon_2 \right)
    \end{align} 
where $\rho=\langle \rho(\epsilon) \rangle$ and 
\begin{align}
    \label{eq:SineKernel}
    K_\textrm{sin}(x,y)= \frac{\sin(\pi(x-y))}{\pi(x-y)},
\end{align}
is the so-called sine-kernel.
However, upon relaxing the condition $ \epsilon \gg \Delta(\epsilon)$ down to the near edge regime $ \epsilon \gtrsim \Delta(\epsilon)$, these results  must be adjusted somewhat:
\begin{itemize}
    \item Eq.~\eqref{eq:R2SineKernel} requires a sufficiently large ensemble of
    levels, such that $ \langle \rho(\epsilon) \rangle$  remains approximately
    constant over the correlation interval $\omega$ centered around $\epsilon$.
    This condition is satisfied far from the edge, $\epsilon  \gg
    \Delta(\epsilon)$. (In a strict sense, Eq.~\eqref{eq:R2SineKernel} holds in the limit $D\to
    \infty$, $\Delta\to 0$ at $\omega / \Delta$ fixed. In sections
    \ref{sec:Dense} and \ref{sec:Sparse} we will discuss how this result changes
    upon approaching the edge of dense or sparse systems.  
    \item The notation $\langle \dots \rangle_\textrm{coll}$ in
    Eq.~\eqref{eq:R2SineKernel} indicates that the background spectral density
    $\rho(\epsilon)$ featuring  in the definition of the dimensionless scaling
    parameter $s$   may be subject to collective fluctuations
    \cite{french1972analysis,Flores:2000ew}. In section \ref{sec:Sparse}, we
    will discuss how for sparse systems the combined inclusion of microscale
    correlations and collective fluctuations determines the statistics of
    spectra down to the smallest energy scales.   
\end{itemize}
Having defined the general setting, we now turn to the specific discussion of 
the two principal universality classes.  
In both classes, the edge has the status of a quantum critical point, with  $
\langle \rho(\epsilon) \rangle$ as a symmetry breaking order parameter, and Eq.
\eqref{eq:SpectralTwoPointDef} as an order parameter correlation function. Much
as in other critical theories, the scaling regime unfolding near the critical point will
be a powerful source of universality. The \emph{difference} between the two
classes has to do with stronger conditions on effective unitary invariance
constraining the dense class, as we will discuss in the following.

\section{Dense systems}
\label{sec:Dense}

We implicitly define the dense universality class  through phenomenological
equivalence to  a Gaussian distributed random matrix Hamiltonian $H$ .
(The condition of Gaussianity can be generalized to other unitarily invariant
distributions, i.e.,  distributions depending on functions of $\tr(H^n)$,
without significant changes to the phenomenology discussed below.) Intuitively,
one expects these `maximally random' ensembles to  model systems whose microscopic Hamiltonian
contains a sufficiently large number of independent matrix
elements, hence the attribute `dense'. (At this point, we remain vague about the
threshold required for a crossover into the dense encoding limit. However, in
section \ref{sec:Numerical} we will investigate this question  for a class of
generalized SYK models with $q$-body interaction, $q\gg 1$.)

In practical terms, the random matrix paradigm implies an ergodicity  condition:
all observables are required to be invariant under unitary transformations in
Hilbert space. In the following, we discuss how this criterion, combined with
the interpretation of the edge as a critical point, essentially fixes the near
edge theory of the dense class. To illustrate this point, the discussion in this
section will be phenomenological, relying only on the above two criteria. (For a
complementary microscopic derivation of the same effective edge theory we refer
to Appendix \ref{sec:MatrixTheoryDetails}.)

Starting from Eq.~\eqref{eq:RhoResolvent} as a definition of our order
parameter, and representing the trace $\tr(G)= \sum_{\mu} G_{\mu \mu}$ as a sum
over the states of an arbitrary Hilbert space basis $|\mu \rangle$,
$\mu=1,\dots,D$, it is evident that we need control over correlation functions
of the structure $\langle G_{\mu \mu}(\epsilon^+)\rangle $ and $ \langle G_{\mu
\mu}(\epsilon_1^\pm) G_{\nu \nu}(\epsilon_2^\pm) \rangle$ where pairwise
summation over $\mu,\nu$ is implicit. 

To describe the symmetry breaking phenomenon in this setting, we 
bundle the
required data in a four-block diagonal operator $\hat G=(\hat \epsilon-H \otimes \mathds{1}_4)^{-1}$, with
\begin{align}
    \label{eq:EnergyMatrixDef}
    \hat \epsilon=\textrm{diag}(\epsilon_1^+,\epsilon_2^-,\epsilon_3^+,\epsilon_4^-),
\end{align}
where $\epsilon^\pm = \epsilon\pm i \delta$. In the limit $\epsilon_1=\dots=
\epsilon_4\equiv \epsilon$, and  for almost all $\epsilon$  (except for those
sitting at the poles of $ H$), this operator is  invariant under similarity transformations,
$\hat G=T \hat G T^{-1}$, $T\in\mathrm{GL}(4)$, up to the
infinitesimal symmetry breaking  $\delta$. Upon averaging over
realizations, we expect a broadening, $\epsilon^\pm \to \epsilon\pm i \gamma$
for energies \emph{inside} the spectral edge, and hence a collapse of the
symmetry to $T \to T^+ \oplus T^-$, where $T^\pm\in \mathrm{GL}(2)$.
Outside the edge, the symmetry remains unbroken. Our goal now is to formulate a
minimal theory for this particular type of symmetry breaking phenomenon. We will
here proceed in a bootstrap manner by postulating this theory on the basis of
symmetry and consistency arguments, to then discuss how it predicts spectral
properties in agreement with those of microscopically defined models.

Assume that our correlation functions can be obtained from a partition
sum
\begin{align}
    \label{eq:PartitionSum}
    \mathcal{Z} (\hat \epsilon)\equiv \int dA \, \exp(-S[A,\hat \epsilon])
\end{align}
by differentiation in the arguments $\hat \epsilon$. In view of the above
symmetry, it is natural to expect that the integration degrees of freedom, $A$,
are \emph{matrices} $A=\{A^{\alpha \beta}\}$, $\alpha,\beta=1,\dots,4$
inheriting $\hat G$'s transformation under the symmetry as $A\to T A T^{-1}$.
There now comes a little twist, namely, we would like the partition sum to be
normalized such  that for $\epsilon_1^+ = \epsilon_3^+$ and
$\epsilon_2^-=\epsilon_4^-$ we have $\mathcal{Z}(\hat{\epsilon} )=1$. This
normalization (which is often realized via the introduction of replica indices)
can be conveniently implemented by `grading' the matrices $A$~\cite{Efetbook}: organizing the
index $\alpha=(a,s)$ into a causal index $s=\pm =\pm1$ and a complementary index
$a=1,2$, we organize the $a$-blocks of the matrix $A$ such that 
\begin{align}
    \label{eq:GradedStructure}
    A=\{A^{ab}\}=\left(\begin{matrix}
        A^{\mathrm{b b}} & A^{\mathrm{b f}}\cr A^{\mathrm{f b}}& A^{\mathrm{f f}}
    \end{matrix} \right),
\end{align}
where $A^\mathrm{b b,f f}$ are $2\times 2$ complex matrices and $A^\mathrm{b f,f
b}$ $2 \times 2$ matrices of Grassmann variables. The `supersymmetry' thus
defined guarantees that in the above limit the functional integral is properly
normalized~\cite{Efetbook}.  In this representation,  the above symmetry is
realized through matrices $T\in \mathrm{GL}(2|2)$, i.e. invertible $4\times 4$
block matrices of the block structure \eqref{eq:GradedStructure}.  (In our
following discussion, normalization by supersymmetry will play only a minor
role. What matters conceptually is that our theory has $4\times 4$ matrices as
effective integration variables.) The matrix structure of $A$ and $G$ suggests a
proportionality $\langle A^{\alpha \alpha}\rangle_A \propto   \langle G_{\mu
\mu}(\epsilon^{\alpha})\rangle_H$ 
where $\epsilon^\alpha$ refers to any of the four energy arguments in
Eq.~\eqref{eq:EnergyMatrixDef} and the averaging on the left/right side of the
equation is over the $A$-functional/microscopic realizations of $H$.

In the following, the matrix variables $A$ will play the role of an effective
order parameter field, conceptually similar to, say, the variable $\phi$ in the
$\phi^4$-theory approach to magnetism.  
Furthering this analogy, we now postulate a minimal action controlling the
fluctuation behavior of these matrices. First, it is natural to postulate
linearity of the action $S$ in the symmetry breaking arguments $\hat{\epsilon}$
(which plays a role akin to a magnetic field in the $\phi^4$-analogy). Noting
that $A\sim \hat G$ carries the effective dimension [energy]$^{-1}$, the
simplest operator satisfying these conditions reads $S_{\hat \epsilon}\equiv c\,
\str(A\hat \epsilon)$ where $c$ is an as yet unspecified constant, and `str' the
trace operation  for graded matrices\cite{Efetbook}, $\textrm{str}(A)\equiv \sum_a
\tr(A^{a a})(-)^{a+1}$. As we are interested in small values of $\epsilon$ near
the edge, we do not include terms of $\mathcal{O}(\epsilon^2)$ in the action. We
next ask which operator might produce symmetry breaking $(\langle A \rangle=0)
\to (\langle A \rangle \not=0)$ as $\epsilon$ changes sign. The simplest
candidate is $c_3 \,\str(A^3)$, giving the $A$-action the form of a cubic
parabola. With this addition the action will have a local minimum or not,
depending on the sign of $\epsilon$, and in this way signal a phase transition.
An optional term of second order $c_2\,\str(A^2)$ can be removed by a shift of
integration variables, $A\to A+\mathrm{const.}$, and higher order terms will be
less relevant in the limit of small $\epsilon$. 

We are thus led to consider the action
\begin{align}
    \label{eq:KontsevichAction}
    S[A,\hat \epsilon]=c\,
\left(\str(A\hat \epsilon)+\frac{1}{3} \str(A^3)\right),
\end{align}
where we used the freedom of rescaling $A\to \lambda A$ to remove one coupling
constant $c_3 \to c/3$.  
The action $S[A]$ defines a supersymmetric version  of the so-called Kontsevich
matrix model~\cite{kontsevich1992}. 

The Kontsevich model provides a universal framework for the description of
spectral correlations near the edges of invariant random matrix Hamiltonians or,
equivalently, random Hamiltonians with dense encoding. Its microscopic
construction detailed in Appendix \ref{sec:MatrixTheoryDetails} implies that from it traces of resolvents
are obtained as
\begin{align}
    \label{eq:CorrelationAExpectation}
    \langle \tr \, G(\epsilon^\alpha)  \rangle &=-\partial_{\epsilon^\alpha} Z(\hat \epsilon),\cr 
    \langle \tr\, G(\epsilon^\alpha) \,\tr\, G(\epsilon^\beta) \rangle&=\partial_{\epsilon^\alpha \epsilon^\beta}Z(\hat \epsilon),
\end{align}
where the derivatives are to be evaluated at the unit-normalized configuration
$\epsilon^+_1=\epsilon^+_3, \epsilon^-_2=\epsilon^-_4$. (For simplicity, we will
mostly consider this configuration, and write $\hat \epsilon = \epsilon +
\frac{\omega}{2}\tau_3$ throughout. The infinitesimal symmetry breaking required
for the differentiations Eq.\eqref{eq:CorrelationAExpectation} is left implicit
in this notation.) In the next two subsections we review how the spectral density and its correlations are obtained from Eqs.~\eqref{eq:CorrelationAExpectation}, followed by integration over the effective $A$-degrees of freedom.

\subsection{Spectral density}

\begin{figure}[t!]
    \centering
    \includegraphics[width=0.8\columnwidth]{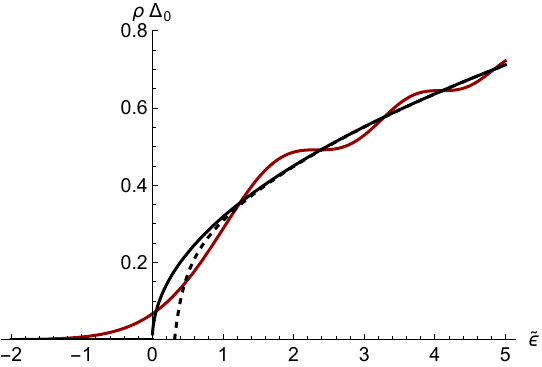}
    \caption{Average spectral density of a dense system near the edge, Eq.~\eqref{eq:SpectralDensityEdgeDense}, expressed as a function of the scaling variable $\tilde \epsilon$. The dashed curve includes the correction Eq.~\eqref{eq:DoSExpansion}.  }
    \label{fig:AiPlot}
\end{figure}

Referring to Ref.~\cite{AltlandSonner21} and Appendix \ref{sec:DenseSpectralDetails} for details, the integration over 
the $A$-matrices yields the average spectral density as
\begin{align}
    \label{eq:SpectralDensityEdgeDense}
    \langle \rho(\epsilon) \rangle = c^{\frac{2}{3}}\left(\tilde \epsilon \textrm{Ai}^2(-\tilde \epsilon)+(\textrm{Ai}'(-\tilde \epsilon))^2\right),\qquad \tilde \epsilon\equiv\epsilon c^{\frac{2}{3}},
\end{align}  
in agreement with the classical result of Ref.~\cite{TracyWidomAiry1993}. Here,
we introduced the dimensionless  variable $\tilde \epsilon$ with scaling factor
$c^{\frac{2}{3}}\sim \Delta_0$.  This is the function schematically shown in
Fig.~\ref{fig:EdgeComparison}, and in quantitative detail in
Fig.~\ref{fig:AiPlot}. Notice how its profile reflects spectral rigidity: the
position of the edge is fixed with (almost) level spacing precision at
$\epsilon=0$, slight deviations showing in an exponentially small tail leaking
beyond the edge. Inside the edge,  oscillatory modulations indicate a tendency
of the levels to order at a separation scale set by the average level spacing.   

An expansion of the Airy functions in large (negative)
arguments
then leads to the asymptotic result $\langle\rho(\epsilon)\rangle\approx
\frac{c}{\pi}\epsilon^{\frac{1}{2}}$. Up to inessential numerical constants, this leads to the identification $\Delta_0 \sim c^{-\frac{2}{3}}$ of the near edge level spacing.  

However, for our purposes, it will be important to go
beyond this approximation and keep track of the first order corrections beyond
it. A somewhat tedious computation based on the inspection of the asymptotic
expansion leads to the result
\begin{align}
    \label{eq:DoSExpansion}
    \langle \rho(\epsilon)\rangle\approx \frac{c^{2 /3}}{\pi} \left( \tilde \epsilon^{1 /2}+ 
    \frac{1}{32 }\tilde \epsilon^{-5 /2}+\textrm{osc.}\right),
\end{align}
where `osc' stands for terms oscillatory as
$\exp(\frac{i4}{3}\epsilon^{\frac{3}{2}}c)$.  In section \ref{sec:Gravity}, we will discuss how
this result can be understood in perturbation theory by diagrammatic methods, or
by topological expansion of the gravitational path integral.

\subsection{Stationary phase analysis and spectral correlations}
\label{subsec:Stationary phase}

\begin{figure}[t!]
    \centering
    \includegraphics[width=1.0\columnwidth]{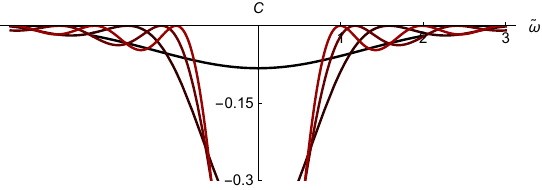}
    \caption{The correlation function Eq.~\eqref{eq:CorrelationFunctionAiry} plotted as a function of the difference coordinate $\tilde \omega=\tilde \epsilon_1 - \tilde\epsilon_2 $ for the center values $\tilde \epsilon = 1,4,7,10$ (color coded). Far from the edge (red) the correlation function approaches the sinusoidal behavior of Eq.~\eqref{eq:R2SineKernel}, close to it (black) it diminishes and loses its oscillatory signatures.}
    \label{fig:AiCorr}
\end{figure}

The integration over the full $A$-space may be extended to an exact computation
of the second moments Eq.~\eqref{eq:CorrelationAExpectation}, cf.
Appendix~\ref{sec:MatrixTheoryDetails}. As a result, one obtains 
\begin{align}
    \label{eq:CorrelationFunctionAiry}
    C(\epsilon_1 ,\epsilon_2 )=- K^2_\textrm{Ai}(-\tilde\epsilon_1,-\tilde\epsilon_2)
\end{align} 
for the connected correlation function (see Fig.~\ref{fig:AiCorr}),
where
\begin{align}
\label{eq:main_AiryKernel}
    K_{\rm Ai}(x,y)
\equiv \frac{{\rm Ai}(x){\rm Ai}'(y)-{\rm Ai}(y){\rm Ai}'(x)}{x-y}
\end{align} 
is the Airy-kernel. For the purposes of our discussion, we will not need this result in
full generality. Instead, we focus on analyzing spectral correlations in a
scaling limit where the center value $\epsilon$ of the energies
$\epsilon_{1,2}^\pm$ is kept fixed while the matrix dimension $D$ is scaled to
infinity. We are thus probing spectral correlations of a sub-ensemble of levels
that can get arbitrarily close to the macroscopic edge but still contains a
large number of levels to obtain universal statistics. In this limit, the exact   
correlation function Eq.~\eqref{eq:CorrelationFunctionAiry} reduces to the
sinusoidal result Eq.~\eqref{eq:R2SineKernel} (cf. Fig.~\ref{fig:AiCorr}) with an average density of states
depending on the center coordinate.

To see in more explicit terms how this comes about, note that in the above  scaling limit, the largeness of $\epsilon\sim D$  invites a stationary
phase approach to the matrix integral. The variation of the action $\delta_A
S[\bar A]=0$ leads to the equation
\begin{align}
    \label{eq:StationaryPhase}
    \hat \epsilon +  \bar A^2=0,
\end{align} 
with the solution 
\begin{align*}
    \bar A = \left\{ \begin{array}{l l}
        (-\hat \epsilon)^{1 /2}\, , & \epsilon <0,\cr 
        -i\hat \epsilon^{1 /2}\tau_3\,,& \epsilon >0,
    \end{array} \right. 
\end{align*}
where in the second line we choose branch of the square root  that
will lead to convergent fluctuation integrals. This solution
makes the interpretation of $\epsilon=0$ as the point of a symmetry breaking
phase transition with $\bar A$ as an order parameter manifest. Above the
transition point, the rotational symmetry of the action in the causal $\pm$
space present in the limit of equal energy arguments, $\omega=0$, is broken, as
indicated by the appearance of the matrix $\tau_3$.

Considering energies inside the symmetry broken phase,   a shift $A\to \bar{A} + A$ defines the effective fluctuation action $S[A+\bar A]=S[\bar{A}
]+S_\textrm{fl}[A]$, with 
\begin{align*}
    S_\textrm{fl}[A]= c \left(-i\,\str\left(\hat \epsilon^{\frac{1}{2}} \tau_3 A^2\right)+ \frac{1}{3}\str(A^3)\right).
\end{align*}  
The largeness of $\epsilon$ implies that generic fluctuations $A$ are gapped by
a factor $c\sim D \sqrt \epsilon \sim \rho(\epsilon)$. This action may now be
applied in different ways to describe the near-edge spectrum.

\paragraph*{Perturbation theory:} A straightforward option is perturbation theory in the parameter $\tilde \epsilon=\epsilon c^{2 /3}$.  We here expand in the cubic nonlinearity, followed by Gaussian integration over $A$ applying Wick's theorem.   The  advantage of this
seemingly redundant approach --- we do know the exact answers, after all ---
is that it provides `semiclassical' clues as to the structure of universal
contributions to the spectral density beyond leading order
$\rho(\epsilon)\sim \epsilon^{1 /2}$. We will discuss this approach, and its
applications in the holographic context in section \ref{sec:Gravity}.

\paragraph*{Double scaling: }
Consider the limit  where the characteristic energy \emph{differences} defined
by the representation $\hat{\epsilon}=\epsilon
\mathds{1}+\frac{\omega}{2}\tau_3$ are small in the sense that $\omega$ is of
the order of a fixed multiple of the characteristic level spacing $\sim
\rho^{-1}(\epsilon)$, and hence of $\mathcal{O}(D^0)$. In this so-called double
scaling limit, generic fluctuations remain gapped in $\sim D$, while the
Goldstone mode fluctuations of the symmetry breaking transition become light. To
see this in explicit terms, note that in the limit of the absence of  explicit symmetry
breaking, $\omega\to 0$, generalizations $\bar A \to T \bar A T^{-1}$ with
rotation matrices $T\in \textrm{U}(2|2)$ are solutions of
Eq.~\eqref{eq:StationaryPhase}\footnote{Here, we do not exploit the full symmetry of the solution space $\textrm{GL}(2|2)\supset \textrm{U}(2|2)$ to obtain a convergent fluctuation integral~\cite{Efetbook}.}.  To leading order in $D^{-1}$, we may thus
approximate the effective action as $S[A]\approx S[T\bar A T^{-1}]\equiv S[Q]$,
where $\bar A$ is the above solution of Eq.~\eqref{eq:StationaryPhase} in the
limit $\omega\to 0$.   Due to the trace invariance, the cubic term now drops
out, and so does the contribution to the energy-vertex proportional to the
homogeneous parameter $\epsilon$.  We are thus left with the soft mode action
\begin{align}
    \label{eq:SoftModeActionDense}
    S[Q]= -i \pi \bar \rho \frac{\omega}{2} \, \str(Q \tau_3),
\end{align}
where $\bar \rho = \bar \rho(\epsilon)=c \epsilon^{\frac{1}{2}}\pi^{-1}$ is the mean field
spectral density.    

Eq.~\eqref{eq:SoftModeActionDense} describes the spectral correlations of
generic ergodic quantum systems with average spectral density
$\rho(\epsilon)$. Referring to  Ref.~\cite{Efetbook} for a detailed discussion, we
note that from this action, the spectral correlation function
Eq.~\eqref{eq:R2SineKernel} is obtained by integration over the manifold of
$Q$-matrices. 

To summarize, in this section we have reviewed how dense systems display a
high level of spectral rigidity, i.e. a near crystalline structure of the
spectrum at the scale of the level spacing. These systems are universally
described by the simple  effective theory Eq.~\eqref{eq:KontsevichAction}. In
fact, one may consider this theory as a \emph{definition} of the dense
universality class near a spectral edge.  This
criterion will become essential in the next section where we discuss ergodic
quantum systems outside this class, which are described by a different effective
theory. 

\section{Sparse systems}
\label{sec:Sparse}

In the following, we contrast the physics of the rigid edge to that of the
softer edge of sparse systems, where the SYK model will be our representative
role model.  
The latter describes $N$ Majorana fermions
$\{\chi_{i}\}$, $[\chi_i,\chi_j]_+=2\delta_{ij}$,  subject  to the interaction
\begin{align}
    \label{eq:SYKHamiltonian}
    H&=\sum_{ijkl}J_{ijkl}\chi_{i}\chi_{j}\chi_{k}\chi_{l}\equiv \sum_\sigma  J_\sigma X_\sigma ,\\
    &\sigma=(i,j,k,l),\qquad X_\sigma =\chi_{i}\chi_{j}\chi_{k}\chi_{l},  \nonumber
\end{align}
where the sum runs over $K\equiv \binom{N}{4}$ four-Majorana index
configurations $\sigma$, and the   interaction coefficients are Gaussian
distributed with variance
\begin{align}
    \label{eq:JVariance}
    \left\langle J_{\sigma}  J_{\sigma'}\right\rangle=
\frac{6J^2}{N^3}\delta_{\sigma \sigma'}.
\end{align}
The model is sparse in the sense that it contains only $\mathcal{O}(N^4)$
statistically independent random coefficients $J_\sigma $ in a Hilbert space of
dimension $D\equiv 2^{\frac{N}{2}}$, 
i.e. the Hilbert space of $N /2$ complex
fermions of conserved fermion parity. Note how $H$ is an `almost diagonal'
operator in that it has matrix elements between fermion occupation number states
$|n\rangle,|m\rangle$ of Hamming distance  $|n-m|\le 4$  (i.e. it changes the occupation of at most four
fermions).   A progressively denser situation  may be realized by
adding to the Hamiltonian higher $q>4$ interaction vertices, likewise with
random coefficients. We will discuss the effect of this generalization in section
\ref{sec:Numerical}.   

The fine grained structure of the SYK spectral density   has been a subject of
intensive
research~\cite{Garcia-Garcia2016,garcia-garciaAnalyticalSpectralDensity2017,berkooz2019}.
Specifically, we know that its average is approximately given by 
\begin{align}
    \label{eq:AverageRhoSYK}
    \langle \rho(E) \rangle\approx c \exp \left( \frac{2\arcsin^2(E /\Gamma)}{\ln \eta} \right),
\end{align}
where 
\begin{align}
    \label{eq:BandWidth}
    \Gamma = 2  \left(\binom{N}{4}\frac{6}{N^3}\right)^{\frac{1}{2}}\frac{J}{1-\eta}\simeq J N
\end{align}
defines the  band width, $c$ is a normalization factor, and $\eta\approx \exp(-2
q^2 /N)$ is a constant approaching the value unity in the limit of large $N$. In
the center of the distribution, $\rho(E)$ is Gaussian. Referring for a more
detailed discussion to Appendix \ref{sec:CordDiagrams}, this difference to the
semicircular spectral density of the dense system, reflects that the
Hamiltonian~\eqref{eq:SYKHamiltonian} contains only polynomially (in $N$) many
operators most of which are commutative. As a consequence, individual many-body
levels assume the form of sums of approximately independent random numbers,
whose statistics is governed by the central limit theorem\cite{French1970}. (By
contrast, the dense system is described by a much larger number of mutually
non-commutative  operators.)

Close to the (say, lower) edge, $ E \approx -\Gamma$, the distribution ceases to be Gaussian, and an near edge expansion of the spectral density leads to the approximation 
\begin{align}
    \label{eq:RhoSYKNearEdge}
    \langle \rho(\epsilon) \rangle\sim  \sinh\left(\frac{2\pi \sqrt{2}}{-\ln \eta}\sqrt{\frac{\epsilon}{\Gamma}}\right)\Theta(\epsilon),
\end{align}    
where $\epsilon =E+ \Gamma$ measures the distance to the edge. 
For $\epsilon\ll \Gamma$, we recover a square root singularity
$\sim \sqrt{\epsilon}$, as for the dense  system. 

These results describe the spectral density to leading order in a $N^{-1}$
expansion, i.e. in a limit in which $N\to \infty$ at a fixed ratio $\epsilon /
\Gamma$. However, beyond this approximation the interplay of two sources of
fluctuations leads to additional deviations from the spectral density of dense systems. We refer to these two as \emph{collective}, and \emph{micro
scale} fluctuations, respectively.

\subsection{Collective spectral fluctuations}
\label{sec:CollectiveSpectralFluctuations}

Choosing a different realization of the SYK Hamiltonian changes a polynomial
number, $K$, of  random parameters in a Hilbert space of exponentially large
dimension, $D$. This turning of relatively few `knobs' leads to large
sample--to--sample fluctuations of the average spectral density via a mechanism  absent
in the dense system \cite{Flores:2000ew,Jia:2019orl,gharibyanOnsetRandomMatrix2018}. 
The same perturbative construction which underlies
Eq.~\eqref{eq:AverageRhoSYK} shows that (cf. Appendix \ref{sec:CordDiagrams})
\begin{align}
    \label{eq:VarDoS}
    \mathrm{var}(\rho(\epsilon))=\frac{ \langle\rho(\epsilon)\rangle^2}{2K}.
\end{align}  

\begin{figure}[t!]
    \centering
    \includegraphics[width=0.8\columnwidth]{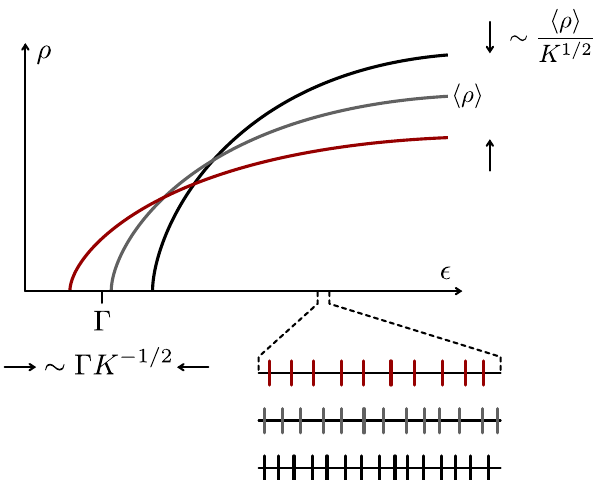}
    \caption{Collective  fluctuations cause a blurring of the spectral edge, fluctuations around the average spectral density, and of the average level spacing.}
    \label{fig:CollectiveFluctuationsSpectrum}
\end{figure}

The  conservation of the total number of levels suggests an interpretation of
these fluctuations in terms of a `breathing mode', i.e. a realization specific
scaling of the spectrum as $\epsilon_\lambda \to \epsilon_\lambda(1+\xi)$
\cite{Jia:2019orl}, and
of the average level spacing, $\Delta \to \Delta (1+\xi)$, see
Fig.~\ref{fig:CollectiveFluctuationsSpectrum}. Assuming a statistical
distribution of the parameter $\xi$, comparison with Eq.~\eqref{eq:VarDoS} shows
that $\mathrm{var}(\xi)=\frac{1}{2K}$, i.e. a scaling of the spectrum
proportional to that of the spectral density. In particular, one expects that
the spectral edge, too, is statistically distributed (also cf. Ref.~\cite{berkooz2019}),    
\begin{align}
    \label{eq:SYKEdgeDistribution}
     -\Gamma \to -\Gamma  + \delta \epsilon_0, \qquad \langle \delta \epsilon_0 \rangle=0,\qquad \textrm{var}(\delta \epsilon_0)=
    \frac{\Gamma^2}{2 K}
    . 
\end{align}
This result implies a softening of the near-edge spectral density as
\begin{align} \label{eq:syk_dos_sinh}
    \langle \rho(\epsilon) \rangle  \approx \left \langle \sinh\left(2\pi \sqrt{2}\sqrt{\frac{\epsilon-\delta\epsilon_0}{\Gamma}}\right)\Theta(\epsilon-\delta\epsilon_0) \right \rangle_{\delta\epsilon_0},
\end{align} 
over the Gaussian distribution specified by Eq.~\eqref{eq:SYKEdgeDistribution}.
In section \ref{sec:Numerical} we  show that this prediction is in excellent agreement with
exact diagonalization. In particular, we expect the integration over the
breathing mode to eradicate  oscillatory signatures in the spectral density.

More generally,  table \ref{tab:1} lists the coarse grained  differences in the
spectral densities of dense and sparse systems, respectively, as caused by
this channel of fluctuations. 
However, before discussing the structure of  the spectrum on the highest resolution
scales,  we need to discuss the complementary microscale correlation channel in the next section. 

\begin{table}[t!]
    \centering\begin{tabular}{|l|c|c|}\hline
        signature & dense & sparse\cr \hline
        collective fluctuations & none & $\mathrm{rms}(\rho)\sim\rho/ \sqrt{K}$
        \cr tail beyond edge & $\mathcal{O}(D^{-1})$
        &$\mathcal{O}(\ln(D)^{-2})$\cr singular power law corrections & yes &
        no\cr near edge oscillatory modulation & yes & no\cr\hline
    \end{tabular}
    \caption{Signatures of the average spectral densities for dense and sparse systems, respectively.}
    \label{tab:1}
\end{table}

\subsection*{Microscale correlations}

For a given realization, we expect the spectrum of a sparse system to show
strong level repulsion as in the dense case, see Fig.
\ref{fig:CollectiveFluctuationsSpectrum} for a qualitative illustration. As we are going to discuss next, this
microscale level structure  is described by the Goldstone modes of the causal
symmetry breaking transition, i.e. the $Q$-modes of our previous discussion, now operating in the sparse context.

Referring to Appendix~\ref{sec:EffectiveSYK} for a microscopic construction, the combined effect of
microscale and collective fluctuations affords a simple representation in terms
of the effective action
\begin{align}
    \label{eq:ActionCollectiveFluctuations}
    S[Q]= \ln  \left \langle \exp\left( -i \pi \rho(\epsilon) \frac{\omega}{2} \, \str(Q \tau_3)\right) \right \rangle_\textrm{coll},
\end{align} 
where the averaging is over the statistically distributed spectral density
$\rho(\epsilon)$ with variance \eqref{eq:VarDoS}. This action describes spectral
correlations between levels $\epsilon\pm \omega$, where we assume $\epsilon \gg
\Delta$. (In view of the large fluctuations, Eq.~\eqref{eq:SYKEdgeDistribution},
there exists no edge that could be approached with a precision set by $\Delta$.) 

There are different ways to extract information from this representation. One
is to integrate over the Gaussian distribution of $\rho$ with variance
Eq.~\eqref{eq:VarDoS}, to obtain the averaged action 
\begin{align}
    \label{eq:ActionCollectiveFluctuationsAveraged}
    S[Q]= -i \pi  \langle \rho(\epsilon) \rangle  \frac{\omega}{2} \, \str(Q \tau_3)
    + 
    \frac{(\pi \omega \langle  \rho(\epsilon)  \rangle)^2}{16K}\,
    (\str(Q \tau_3))^2. 
\end{align} 
One may then obtain the spectral correlation functions by integration over the
$Q$-matrices. Due to the high symmetry of the induced quadratic weight, this
procedure is not  more difficult than the integration over
Eq.~\eqref{eq:SoftModeActionDense}. However, an even more straightforward strategy is to first integrate over $Q$ in
Eq.~\eqref{eq:ActionCollectiveFluctuations}, i.e. to compute spectral
correlations for a given realization $\rho(\epsilon)$. As a result, one obtains
Eq.~\eqref{eq:R2SineKernel}, which in a second step is averaged over
$\rho(\epsilon)$. 

Either way, the result of these computations, is best discussed in the language
of the spectral form factor, i.e. the temporal Fourier transform 
\begin{align}
    \label{eq:SpectralFormFactorDef}
    K(\tau) = \int ds \, e^{is \tau} C(s)
\end{align}
of the spectral correlation function expressed in the energy-like dimensionless
variable, $s=\pi \omega \rho$. For Eq.~\eqref{eq:R2SineKernel}, we obtain the
familiar result
\begin{align*}
    K(\tau)=\tau \Theta(1-\tau)+\Theta(\tau-1),
\end{align*}
comprising a linear ramp for $\tau<1$ followed by a plateau for larger times. 

We may now define the physical form factor through $K_\mathrm{phys}(\tau)\equiv
\pi \bar \rho \langle \tilde K(t) \rangle$, where $\tilde K(t)\equiv \int d
\omega \exp(i \omega t) \langle C_\rho(\omega) \rangle$ is the real time Fourier
transform of the correlation function \eqref{eq:SpectralTwoPointDef} and the notation emphasizes that
$C_\rho(\omega)=C(\epsilon_1,\epsilon_2)$ depends on the difference
$\omega=\epsilon_1-\epsilon_2$, and on the center energy through
$\rho=\rho(\epsilon)$. In this way, we obtain
\begin{align}
    \label{eq:FormFactorRounding}
    K_\mathrm{phys}(\tau)= \left \langle \frac{ \langle \rho\rangle }{\rho} K\left(\tau \frac{ \langle \rho\rangle }{\rho} \right) \right \rangle.
\end{align} 
The distribution \eqref{eq:VarDoS} implies that $ \langle \rho\rangle  /\rho $
is a random variable with average unity, and variance $\mathrm{var} ( \langle
\rho\rangle  / \rho)= 1 /(2K)$. The most visible influence of the average over
the distribution is a rounding, over scales $\sim K^{-1 /2}$, of the corner
singularity of $K(\tau)$ at $\tau=1$, i.e. at physical time scales of the order
of the Heisenberg time $\sim \Delta^{-1}$. 

The principle here is that collective
fluctuations of the spectral density over large scales let the average level
spacing, and hence the oscillation period of the spectral correlation functions
at the smallest scales fluctuate as well. The effects of this locking mechanism are  witnessed, e.g., by the
rounding of the spectral form factor.

\section{Numerical analysis}
\label{sec:Numerical}

In this section we compare the results discussed above to exact diagonalization.
We begin with a discussion of the standard SYK model, and then consider a
generalized variant where the interaction vertex couples $q> 4$ 
Majorana fermions.

\begin{figure*}
    \centering
    \includegraphics[width=.24\textwidth]{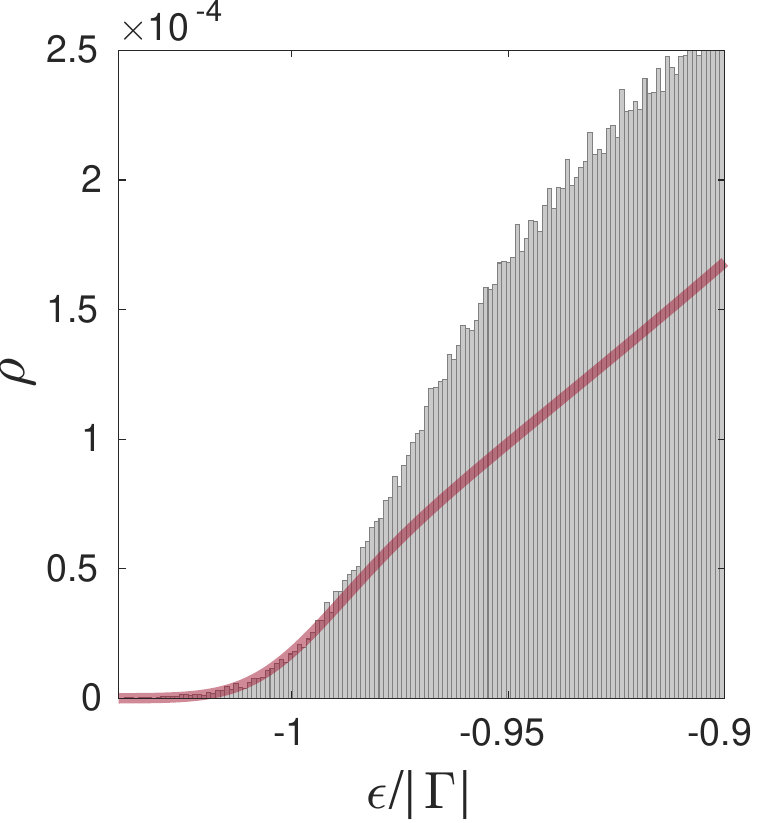}
    \includegraphics[width=.24\textwidth]{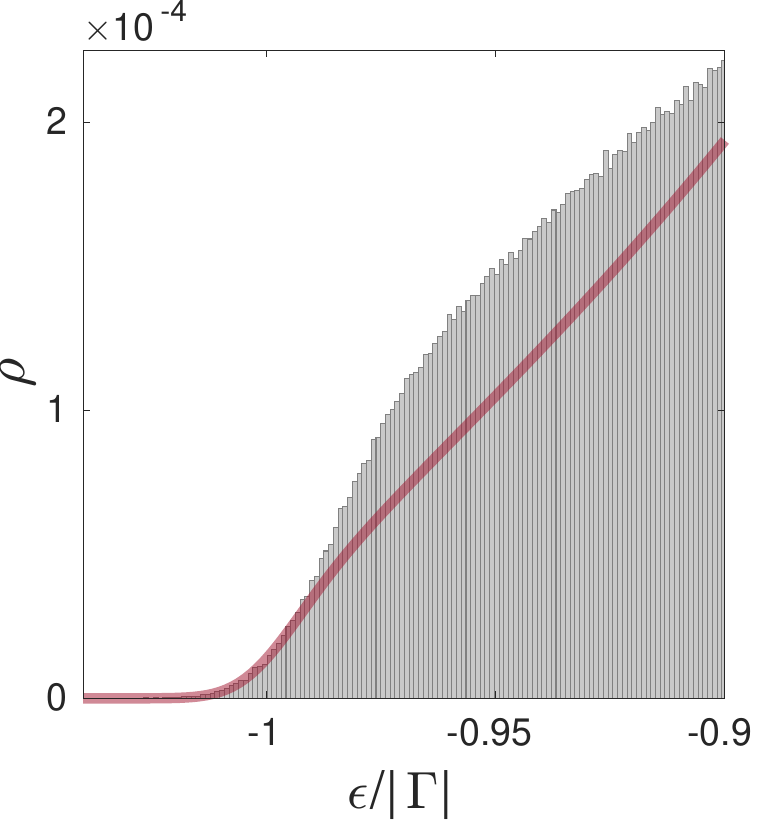}
    \includegraphics[width=.24\textwidth]{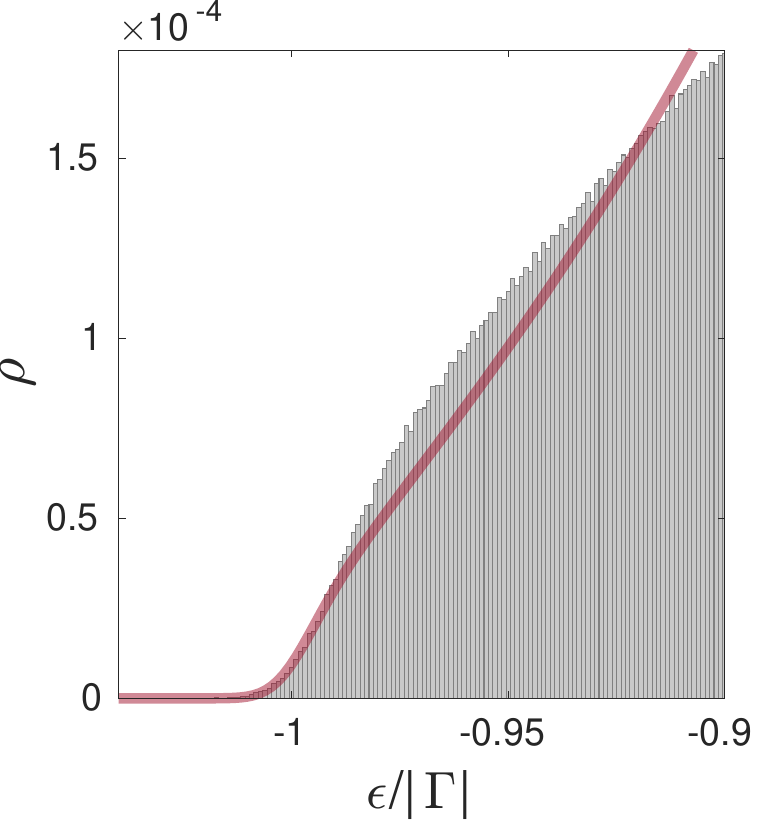}
    \includegraphics[width=.24\textwidth]{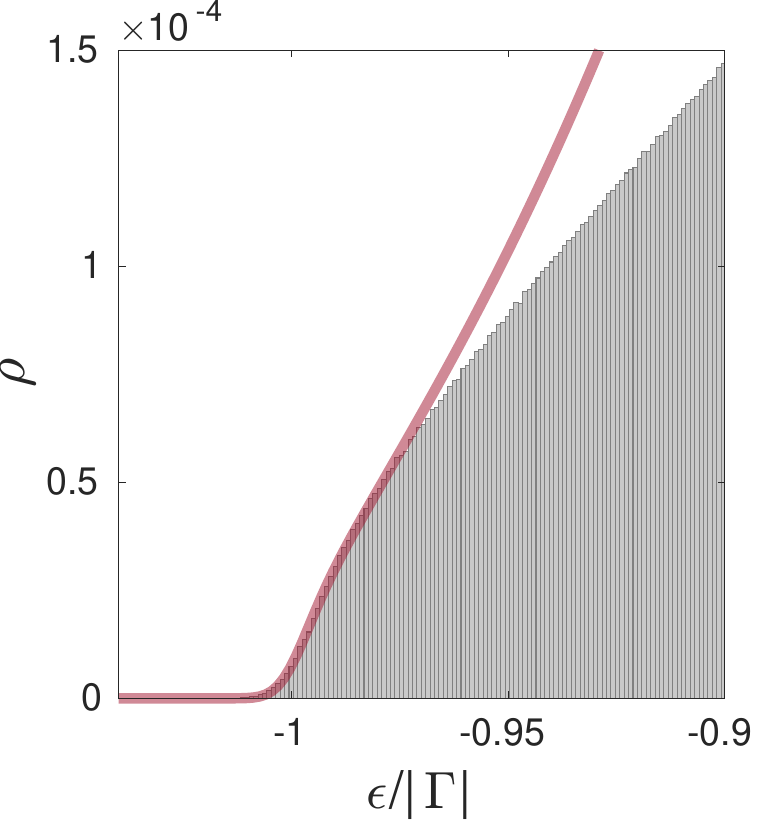}
    \caption{\label{fig:sinh} Left to right: Near edge spectral density of the SYK model with $q=4$ and $N=18,\; 22,\; 26,\; 30$. We obtain a progressively better fit to 
     Eq.~\eqref{eq:syk_dos_sinh} in terms of two fitting parameters $\Gamma$ and the normalization prefactor  (red curves). After removal of 
    fluctuations $\delta \epsilon_0$ (see Eq.~\eqref{eq:SYKEdgeDistribution}), the $\sinh$ profile  describes the spectral density for $N\ge
    30$. The number of realizations here are $50,000$ for $N=18,\;22$ and $10,000$ for $N=26,\;30$, respectively, and fitting parameters for increasing $N$  are $\Gamma=-3.62,\;-4.28,\;-4.96,\;-5.62$.
    }
    \end{figure*}

\subsection*{SYK Model}
Fig.~\ref{fig:sinh} shows the near edge spectral density of SYK models with
$N=18,\;22,\;26$ and $30$ Majorana fermions. (We
increase in steps of $4$ to keep the  model in the unitary symmetry
class~\cite{You2017}.)  The red curves are fits to
Eq.~\ref{eq:syk_dos_sinh} in terms of two parameters, $\Gamma$ and the overall
prefactor omitted in that formula. For increasing $N$, the agreement becomes excellent and extends
deeper into the spectrum. We have also checked that for large $N\ge 30$ and  after removal of the
collective fluctuations of the edge position Eq.~\eqref{eq:SYKEdgeDistribution},
the  distribution of individual systems follows the
$\sinh$-profile of Eq.~\eqref{eq:syk_dos_sinh}, see also Ref. \cite{garcia-garciaAnalyticalSpectralDensity2017}.   
   \begin{figure}[b!]
\centering
\includegraphics[width=0.45\textwidth]{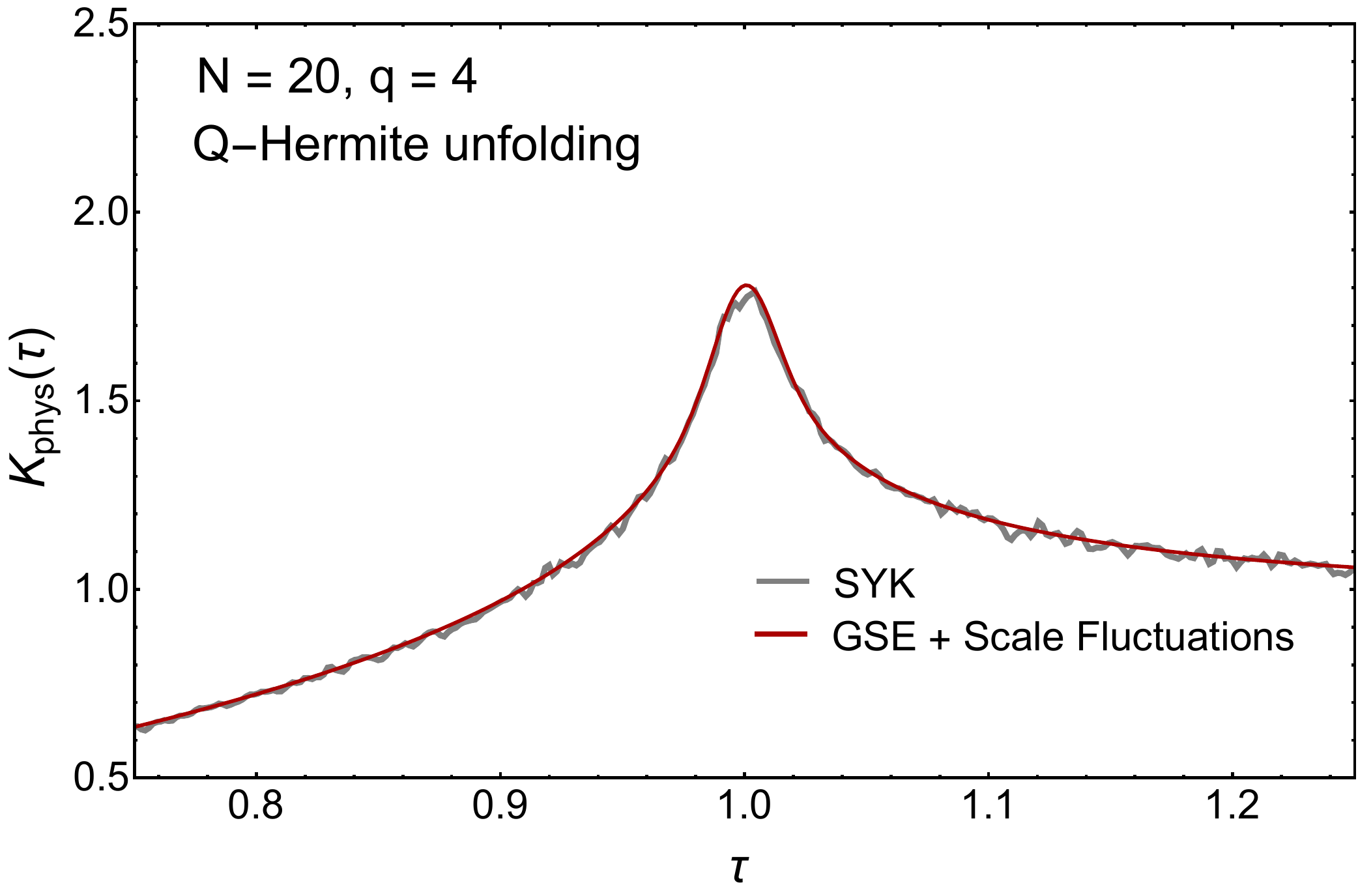}\\
\includegraphics[width=0.45\textwidth]{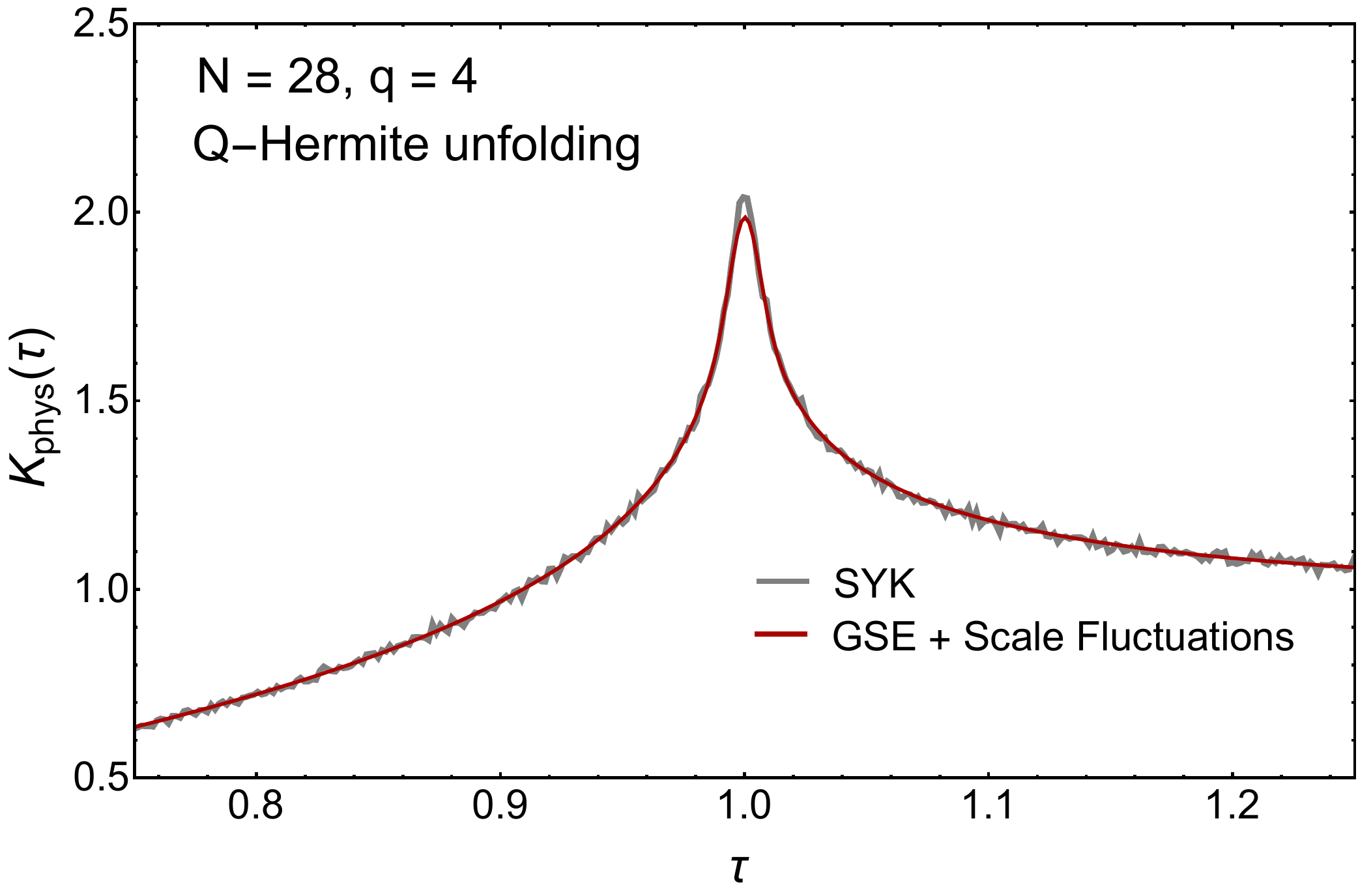}
\caption{Average connected spectral form factor for $q=4$ SYK spectra (gray curves) for $N=20$ (upper) and $N=28$ 
which are
in the GSE universality class. The results are compared to the analytical GSE result with scale that fluctuates according to a Gaussian
with unit average and variance determined by  moments of the SYK Hamiltonian. The number of realizations for
$N =20$ and $N=28$ is 10,000 and 15,0000, respectively. 
\label{fig:form}
}
\end{figure}

\begin{figure*}
    \centering
    \includegraphics[width=.4\textwidth]{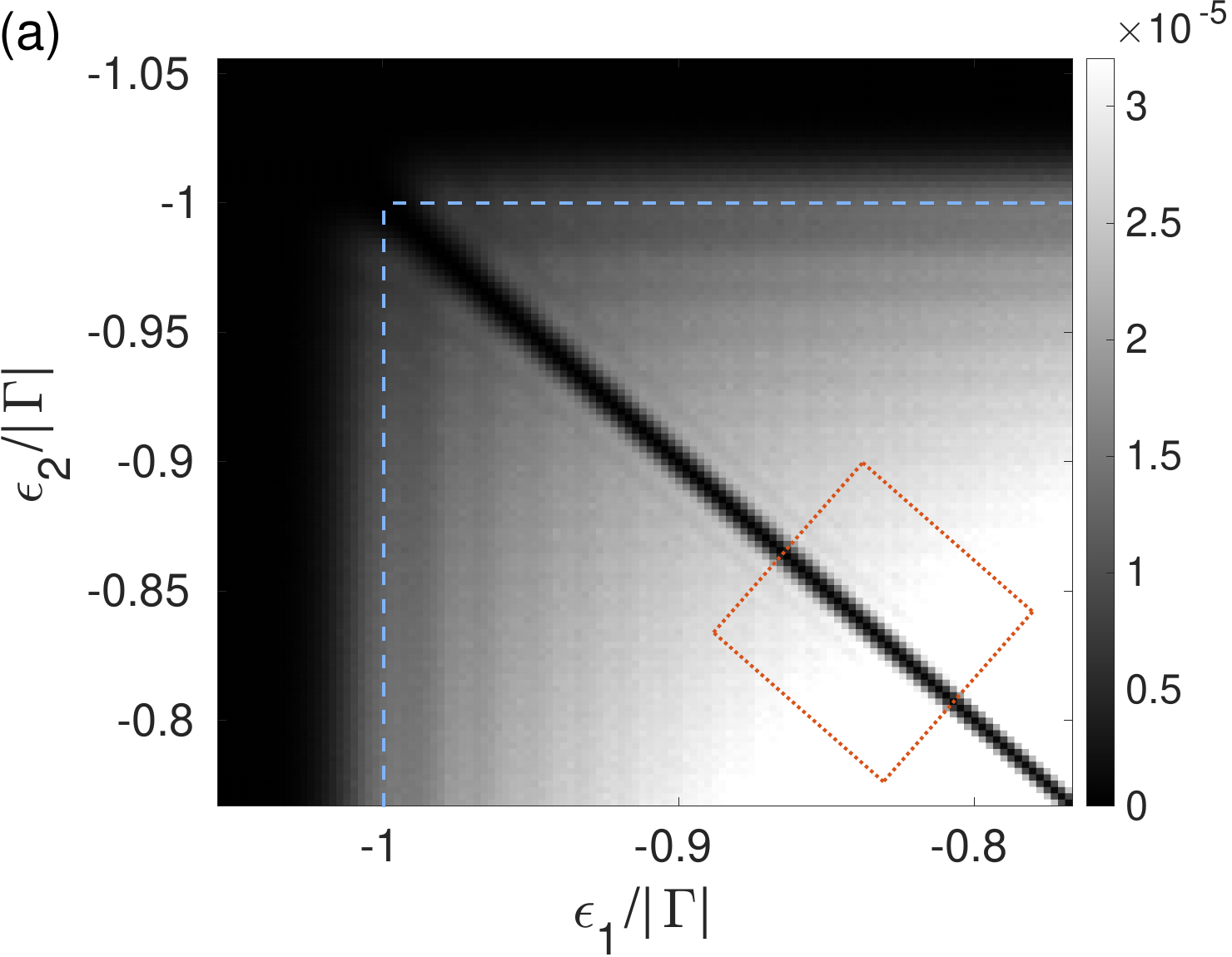}
    \includegraphics[width=.4\textwidth]{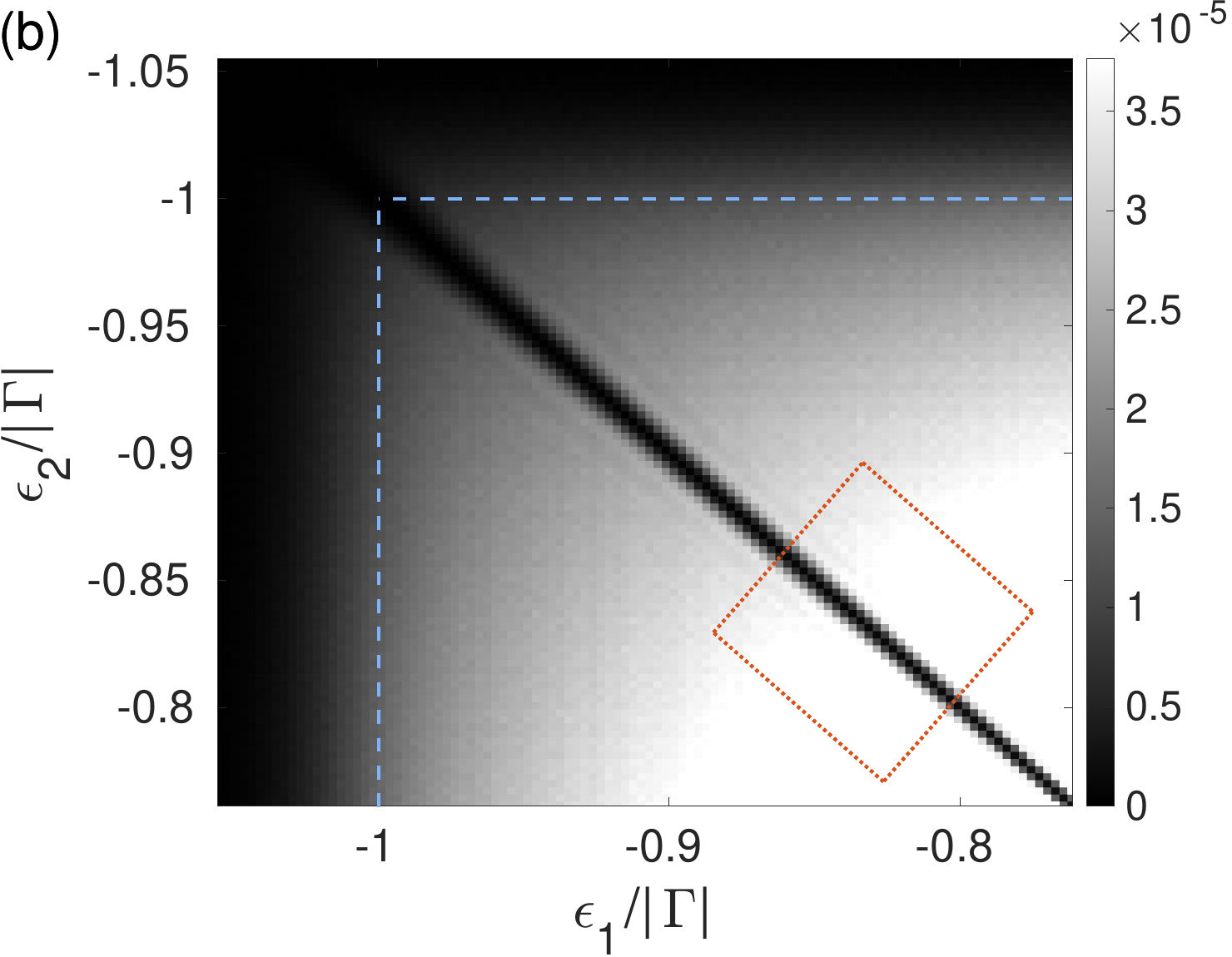}
    \includegraphics[width=.3\textwidth]{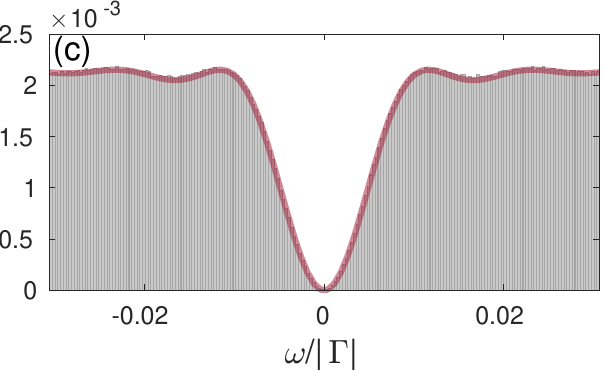}
    \includegraphics[width=.3\textwidth]{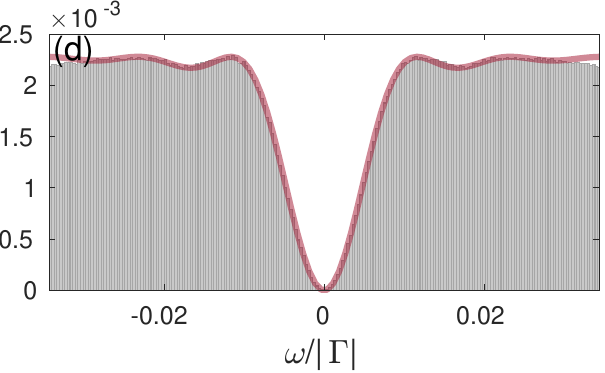}
    \caption{\label{fig:dos_correl} 
    Level-level correlation function\textbf{}
    $\langle \rho(\epsilon_1)\rho(\epsilon_2) \rangle$  for a random matrix ensemble (a)
    and the SYK model with $q=4$, both acting on Hilbert-spaces of dimension $D=2^8$~\cite{footnote1}.
        Notice the oscillations of the average spectral density in the RMT case.
        Panels (c) (RMT) and (d) (SYK model) show the corresponding spectral correlations as a function of the relative energy $\omega=\epsilon_1-\epsilon_2$,
        averaged along the center energy $\epsilon=(\epsilon_1+\epsilon_2)/2$. In both cases the average over $\epsilon$ 
        leads back to the universal result Eq.~\eqref{eq:R2SineKernel}.}
    \end{figure*}

    Figure \ref{fig:dos_correl} shows the spectral correlations $\langle
    \rho(\epsilon+\omega /2) \rho(\epsilon - \omega /2 ) \rangle_c$  near the
    edge, i.e. the correlation function \eqref{eq:SpectralTwoPointDef} without
    the normalizing factor $\Delta^2(\epsilon)$. Panel (a) shows data in a gray
    scale representation for an ensemble of random matrices of dimension $D=2^8$
    and (b) for a SYK model with $N=18$, both averaged over $2\times 10^5$
    realizations.  Upon approaching the edge, the spectral density, and hence its moments, decrease 
    as indicated by the darkening of the color. The main difference between the
    two correlation profiles is the ripple pattern visible in the RMT case, (a).
    It is a consequence of the oscillatory modulations of the near edge spectral
    density, Eq.~\eqref{eq:main_AiryKernel}, which extends to the unnormalized
    correlation function. Another way of stating the same is that the RMT
    correlation function is given by Eq.~\eqref{eq:CorrelationFunctionAiry},
    which only for energies well inside the spectrum, $\epsilon_i\gg \Delta_0$
    asymptotes to the sine kernel correlation function
    Eq.~\eqref{eq:R2SineKernel}. Panels (c) and (d) compound the spectral
    correlations averaged over the center coordinates within the regions
    indicated by dotted parallelograms in (a) and (b). The data show that both
    the SYK model and a reference random matrix model retain their sine kernel
    microscale correlations Eq.~\eqref{eq:R2SineKernel} in the near edge region.

As discussed in the previous section, collective fluctuations result in a rounding of the spectral form factor at the Heisenberg time.
This effect is most pronounced for the Gaussian Symplectic Ensemble (GSE) where the form factor diverges at this point. In Fig. \ref{fig:form}, we
illustrate this for the connected spectral form factor of the $q=4$ SYK model for $N= 20$ (upper) and 
$N=28$ (lower). The gray solid curves represent the results for the unfolded 
SYK spectra. They have been unfolded by means of the exact Q-hermite spectral density corrected 
by an  even 8th order Hermite polynomial with coefficients fitted to the ensemble averaged spectral 
density.
This can be justified from the observation that the spectral form factor is local in the sense that 
only close pairs of eigenvalues result in significant contributions.
The exact data are compared to the analytical result \eqref{eq:SpectralFormFactorDef} with $\mathrm{var} ( \langle \rho\rangle  / \rho)= 1
/(2K)$ and $K(x)$ equal to the spectral form  factor for the GSE. The agreement is excellent.

    \begin{figure*}
        \centering
        \includegraphics[width=0.24\textwidth]{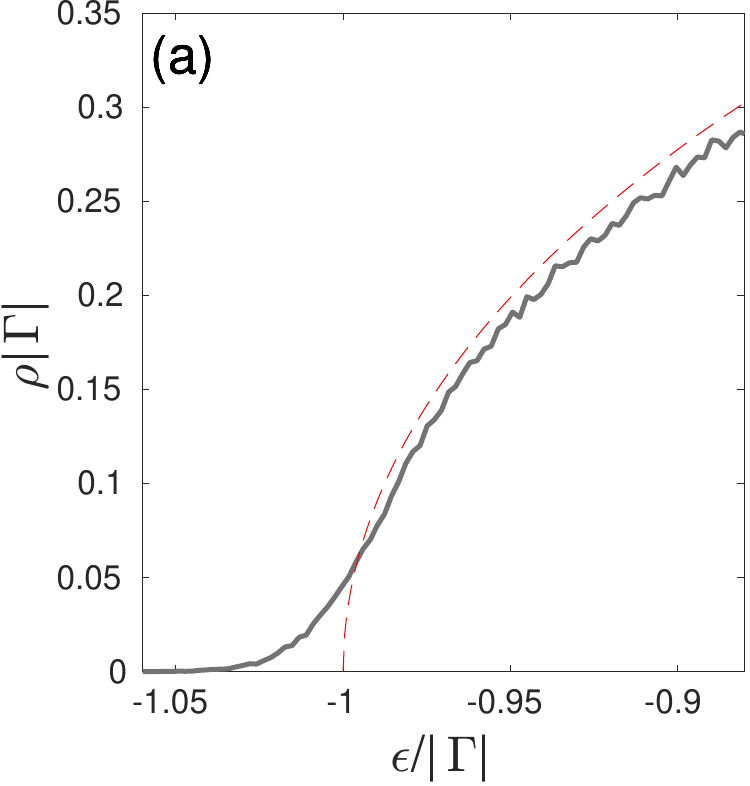}
        \includegraphics[width=0.24\textwidth]{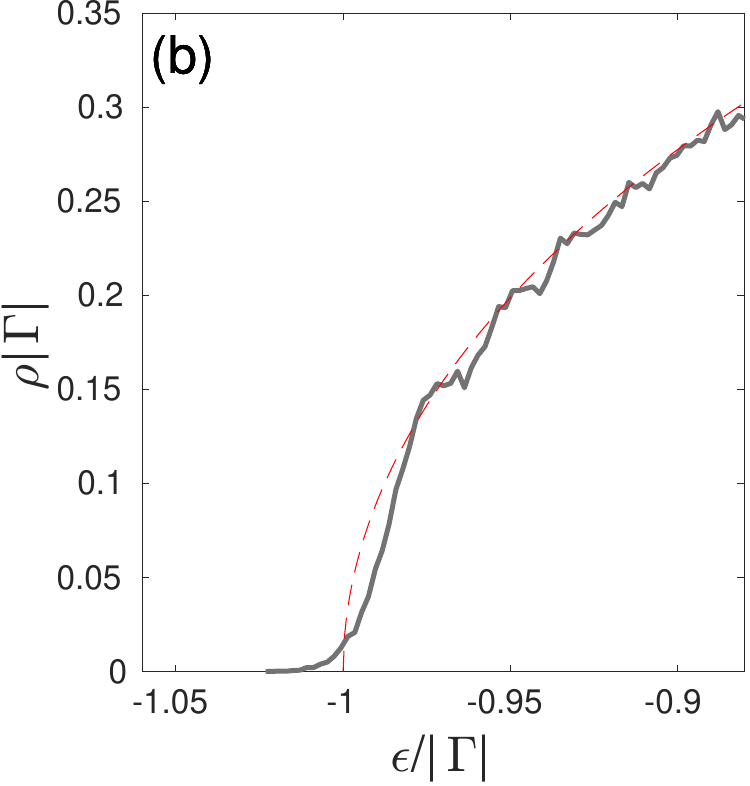}
        \includegraphics[width=0.24\textwidth]{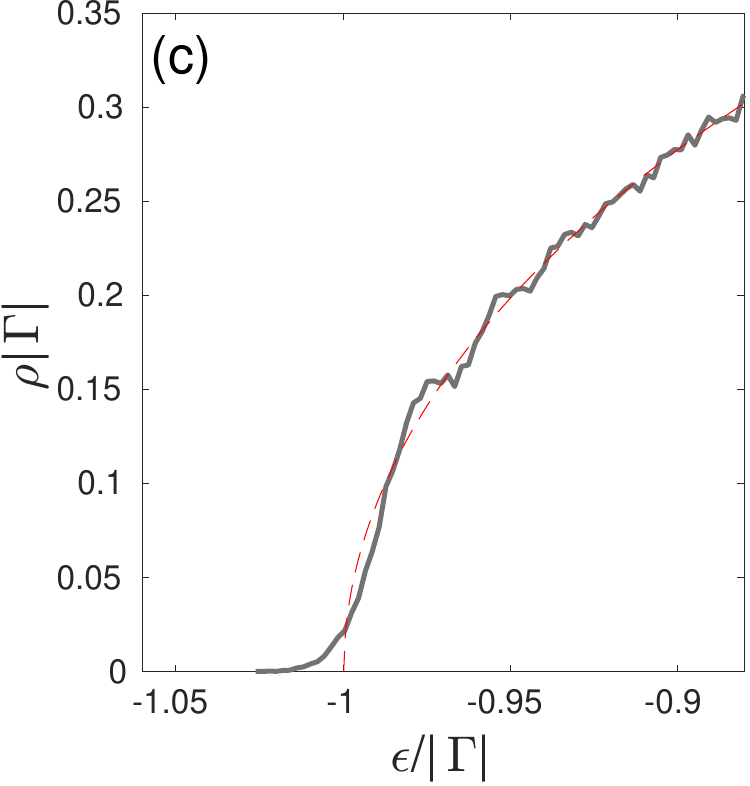}
        \includegraphics[width=0.24\textwidth]{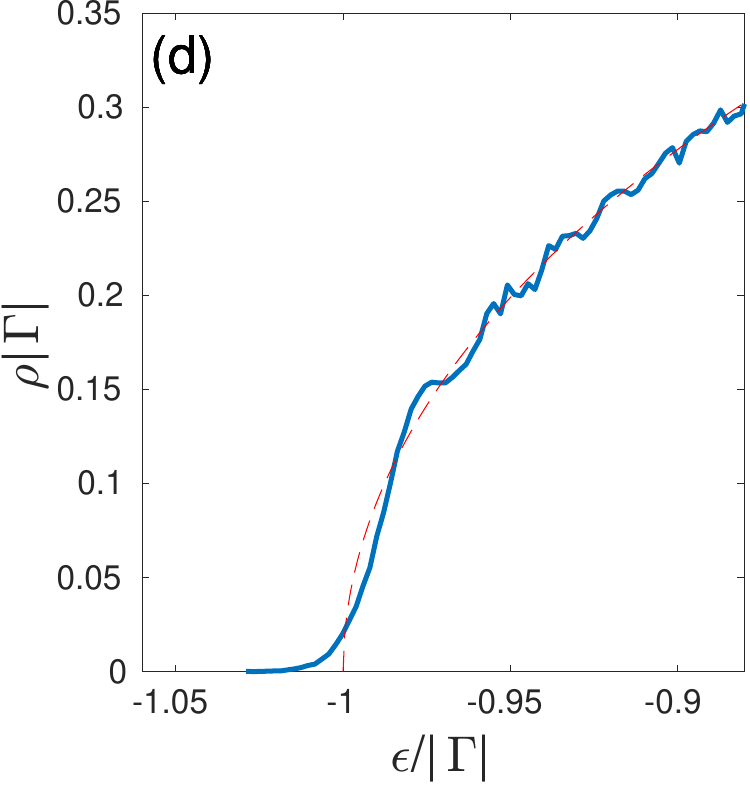}
        \includegraphics[width=0.24\textwidth]{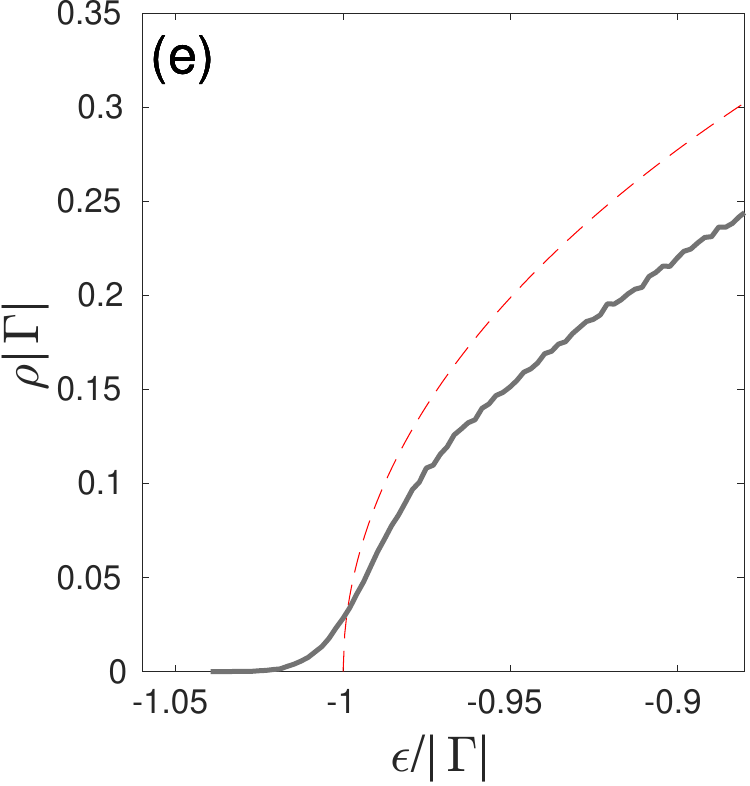}
        \includegraphics[width=0.24\textwidth]{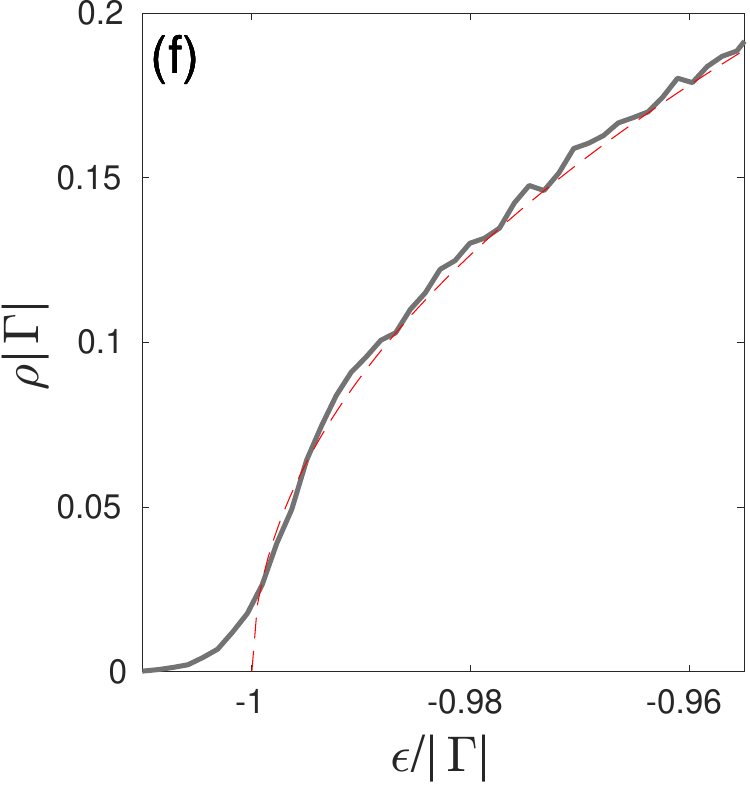}
        \includegraphics[width=0.24\textwidth]{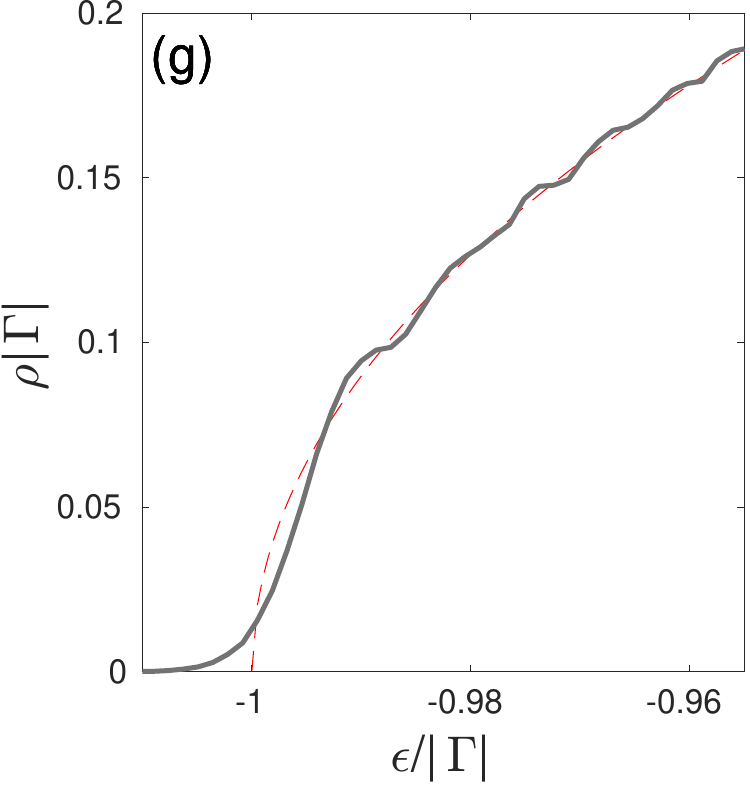}
        \includegraphics[width=0.24\textwidth]{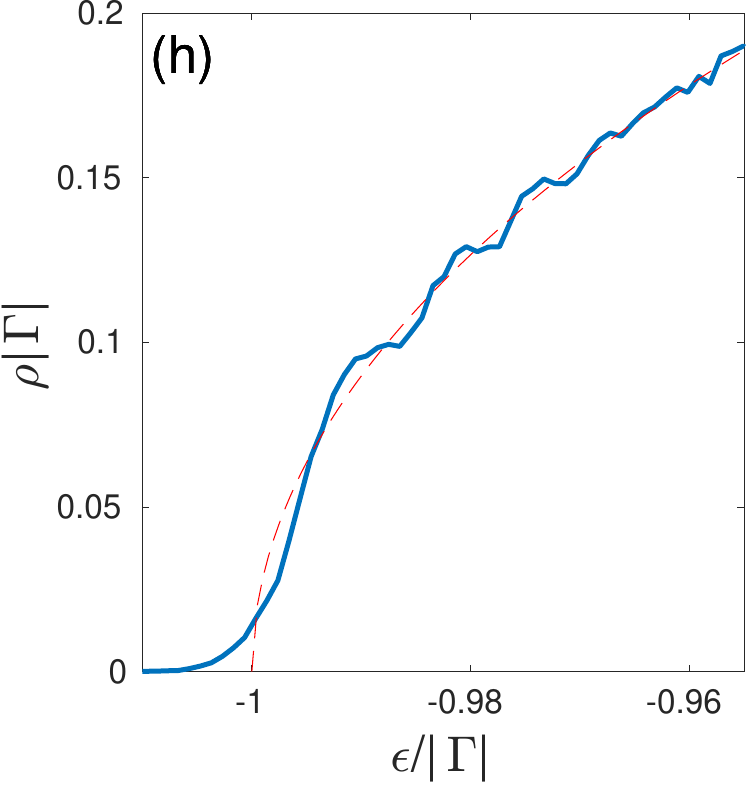}
        \caption{\label{fig:num}  Crossover from sparse to dense spectral statistics probed via the average spectral density of SYK models with $q=4,\;6,\;8$-Majorana interaction for $N=18$ (panels (a-c)) and $N=22$ (panels (e-g)).  The guide
        lines (red dotted) are generated by fitting a semi-circle to the DoS of random
        matrices of dimensions  
            $D=2^8$ and $D=2^{10}$, respectively. The 
            formation of the spectral edge
        oscillation is clearly seen as $q^2/N$ increases for both sizes $N=18,\;22$. 
        Each plot was generated by averaging over an ensemble of 50,000 realizations. 
        }
        \end{figure*}

\subsection*{Generalized SYK Model}

An increasing of the number of random parameters in a many-body system  will
eventually induce a crossover between the spectral signatures of sparse and
dense systems. To realize this phenomenon, we  consider a variant of the SYK
Hamiltonian \eqref{eq:SYKHamiltonian}
\begin{align}
    H = \sum_{i_1<i_2<\cdots<i_q} J_{i_1,\cdots i_q} \chi_{i_1}\chi_{i_2}\cdots \chi_{i_q}, 
\end{align}
containing products of a variable number $q\geq 4$ of Majorana operators. As
before, we define the Gaussian variance of the matrix elements through
Eq.~\eqref{eq:JVariance}, with $\sigma=(i_1,\dots,i_q)$. The spectral density of
the generalized model can be
calculated~\cite{garcia-garciaAnalyticalSpectralDensity2017} by the same methods
as in the standard model (cf. Appendix ~\ref{sec:CordDiagrams}), and continues
to be given by Eq.~\eqref{eq:syk_dos_sinh} with, however, the constant  $\eta$
now generalized to 
\begin{align*}
    \eta=\binom{N}{q}^{-1}\sum_{r=0}^q(-1)^{q+r}\binom{q}{r}\binom{N-r}{N-q}\stackrel{N\gg q}\approx (-1)^qe^{- \frac{2q^2}{N}}.
\end{align*} 
 Figure \ref{fig:num} shows the  averaged DoS of SYK models with increasing
 connectivity $q=4,\;6,\;8$ for $N=18$ (panels (a-c)) and $N=22$ (panels (e-g)).
 We observe a gradual stiffening of the edge and the onset of oscillations which
 for $N=18$ and $q=8$ lead to a profile similar to that of a random matrix
 (panels (d) and (h) for comparison). The onset of the spectrum given by
 $E_0=4\sigma/(1-\eta)$ (with $\sigma$ the width of the spectrum) also
 approaches the RMT value of $2\sigma$. On general grounds, one expects the
 crossover from sparse to dense spectral correlations to be driven by the
 parameter $\lambda \sim q^2 /N$. (The constant of proportionality is of
 $\mathcal{O}(1)$ and  depends on choice of the variance of the coefficients
 $J_\sigma$.)

\section{Application: Two-dimensional holography}
\label{sec:Gravity}

While it had been known for a number of years that low-dimensional quantum
gravity has a close connection with matrix integrals \cite{Ginsparg:1993is}, it
has only been appreciated more recently, that the matrix integrals occurring in
quantum gravity in fact may have their origin in chaotic dynamics. Early work
mainly employed  matrix-model machinery as an elegant and efficient tool to
define a path integral of discrete triangulations of two-dimensional
`universes'. However, recently it has become increasingly clear that the random
matrices occurring in these constructions in fact have an interpretation as
boundary Hamiltonians, in the sense of holographic duality
\cite{Saad2019}. 

The prototypical example of such a correspondence between a two-dimensional
theory of gravity and random-matrix theory is the so-called JT-theory of
gravity, which is perturbatively\footnote{The correspondence is perturbative in that it has been established order by  an expansion equivalent to the above $\tilde \epsilon$-expansion. For its non-perturbative completion in terms of topological string theory, see Ref.~\cite{Post:2022dfi,altlandQuantumChaos2D2023}.} dual to the `SSS matrix' model \cite{Saad2019}. The two-dimensional theory of
gravity in question has a semiclassical description starting with the action
\begin{equation}\label{eq.JTAction}
S_{\rm JT} = -S_0 \chi({\cal M}) -\frac{1}{2} \int_{\cal M} \sqrt{g} \phi\left(R + 2 \right) - S_\textrm{bdy}\,,
\end{equation}
 where $g$ is the metric,
and $R$ the associated scalar curvature. The scalar field $\phi$ is the so-called dilaton, which is present as an extra degree of freedom in JT gravity, and is best understood as a remnant of higher-dimensional gravitational degrees of freedom if viewing JT gravity as a dimensional reduction from higher dimensions \cite{Mertens:2022irh}. The coefficient  $S_0$ defines the
coupling to the Euler number $\chi\left( {\cal M} \right)$ of the manifold
${\cal M}$,  the notation emphasizing its identification with an entropy.
(Within the framework of the holographic correspondence, $S_0$ is the extensive
ground-state entropy of the SYK model.) Finally, the boundary action $S_{\rm bdy} \equiv
\int_{\partial {\cal M}} du \left({\cal K}-1  
\right)$ involves the extrinsic curvature ${\cal K}$ of the boundary $\partial
{\cal M}$, whose details may be found for example in
\cite{Saad2019,Post:2022dfi}. It ensures that the action \eqref{eq.JTAction}
has a well-defined variational principle, as well as a (renormalized)
bulk-boundary dictionary.

The duality between \eqref{eq.JTAction} and a matrix integral of the form
\begin{equation}
    \int dH e^{-V_{\rm SSS}(H)}
\end{equation}
has the status of a holographic duality, similar in spirit to the duality
between ${\cal N}=4$ supersymmetric Yang-Mills theory and type II string theory
on AdS$_5 \times$S$^5$ \cite{Maldacena:1997re} with one major departure: while
the original examples propose a duality between a fixed individual field theory,
namely ${\cal N}=4$ SYM, and a fixed individual bulk theory, namely type IIB
superstrings, the JT/SSS duality in fact juxtaposes a (seemingly) fixed
individual bulk theory (JT gravity) with an {\it ensemble} of boundary theories,
the SSS matrix model with the interpretation that the matrix integral is in fact
over boundary Hamiltonians. This has spurred much activity attempting to
reconcile these two radically different examples of holographic duality,
motivated by the possible implication that gravity might microscopically be
described by an average over theories, rather than any specific theory. Without
entering into the details, which may be found e.g. in
\cite{Saad2019,Blommaert:2019wfy,Marolf:2020xie,AltlandSonner21,Saad:2021rcu,Blommaert:2021fob,Altland:2022xqx}, we
may simply summarize the situation by posing the question ``is gravity dual to
an ensemble?''. 
One way to ease this tension between low- and high-dimensional
incarnations of the holographic principle, might be an interpretation of the
former as lower-dimensional reductions of higher-dimensional parent theories by
integration over hidden degrees of freedom. This view is supported, e.g., by the
explicit construction of Ref.~\cite{Post:2022dfi}, representing two-dimensional
JT gravity as a reduction from a six-dimensional (Calabi-Yau) theory in a manner
resembling an ensemble average.

Another possible way to reconcile these two points of
view would be to exhibit a fixed quantum mechanical theory whose chaotic
behavior is such that it is described by, say, the SSS matrix model -- or indeed
any other example of a matrix model with a gravitational dual in (bulk)
dimension two or three (see \cite{Belin:2020hea,Chandra:2022bqq,Belin:2023efa} for 3D examples) -- together with a suitable coarse-graining procedure for low-energy gravity. For this purpose the dense-encoding properties we have
proposed above give a new discriminatory criterion to distinguish potential
quantum chaotic theories that could give rise to the required averaged physics.


 ``Natural'' candidates for quantum boundary theories in low-dimensional holographic duality would appear to be many-body
models subject to few-body random interactions, such as models in the SYK
family. Here the issue is  made more pressing by the fact that low-dimensional holography unfolds close to the
ground states of the  theories in question, i.e. precisely the region where the
differences between dense and sparse become most acute.

Before commenting further on this point, let us thus briefly
review how JT gravity produces results indicative of such densely encoded random systems. This means that we need to understand how spectral properties are formulated in a gravitational language. The
quickest way to see spectral statistics expressed in terms of gravitational quantities is
via the calculation of multi-resolvent correlators, which are encoded in certain geometries contributing to the path integral with action \eqref{eq.JTAction} above, following the original treatment in \cite{Saad2019}. In fact, it is more natural
in holographic duality to instead work with the Laplace transform and compute the
partition function $Z(\beta)$ and its generalization to the multi-resolvent case. We are therefore interested in computing quantities of the type
\begin{equation}
Z(\beta_1) \cdots Z(\beta_n) \Bigr|_{\rm JT}\,,
\end{equation}
given by spacetimes with $n$ boundary circles of length $\beta_1\,\ldots, \beta_n$, potentially connected to each other by the bulk geometry.
In this way, the density of states is related to the JT path integral over surfaces with one boundary, but arbitrary bulk topology
\begin{equation}\label{eq:JTspectralExpansion}
    Z(\beta)\Bigr|_{\rm JT} = \,\,\,\includegraphics[width=0.3\textwidth, valign=c]{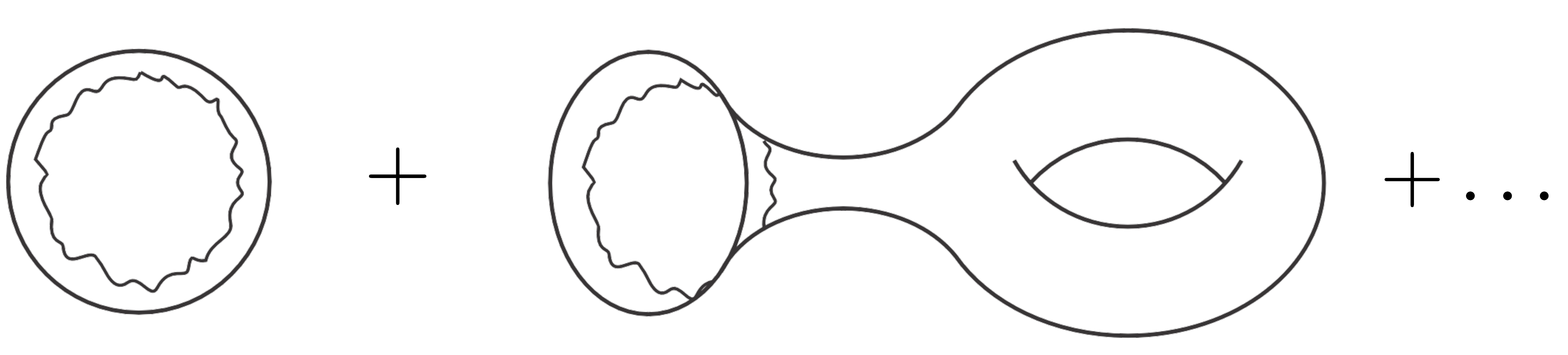}\,
\end{equation} 
Similarly, we may deduce the two-level correlation function from the path
integral over two-boundary geometries in JT gravity
\begin{align*}
    &Z(\beta_1) Z(\beta_2) \Bigr|_{\rm JT} = \cr 
    &\quad \includegraphics[width=0.45\textwidth, valign=c]{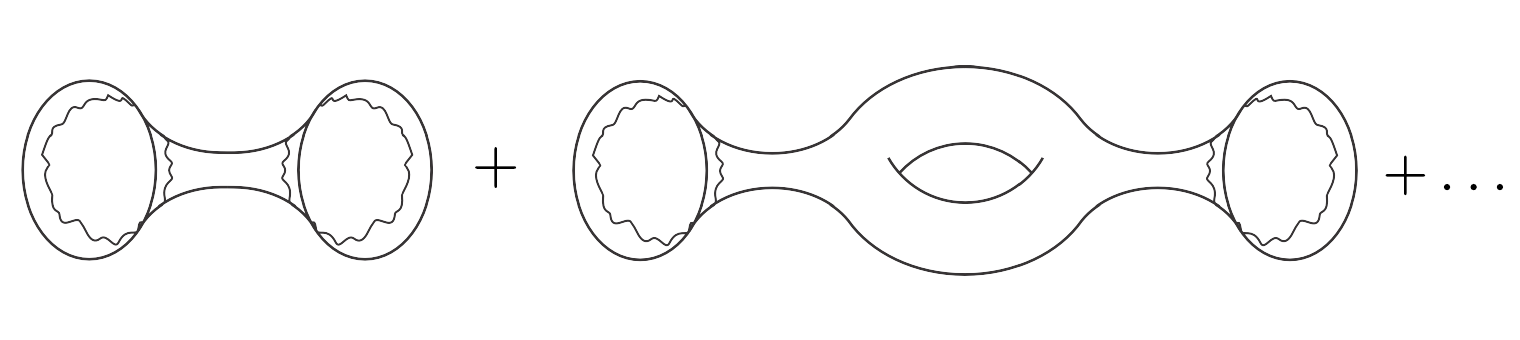}
\end{align*}    
Returning to the spectral density we obtain the leading answer from the disk
partition function
\begin{equation}
    Z_{\rm Schw}^{\rm disk}(\beta) = \frac{e^{\frac{\pi^2}{\beta}}}{\sqrt{16\pi \alpha^3 \beta^3}} \,,
\end{equation}
resulting in
\begin{equation}\label{eq.JTspectralDensity}
    \rho_0 (E) =\frac{e^{S_0}}{(2\pi)^2}\sinh\left( 2\pi \sqrt{E} \right)\,,
\end{equation}
i.e. an expression equal to that of \emph{average} spectral density of the
SYK model Eq.~\eqref{eq:RhoSYKNearEdge} upon identifying the present
dimensionless $E$ as the SYK energy measured in units of the band width.
However, a closer look reveals differences rooted in the fact that JT is a dense
system, while SYK is sparse. These difference emerge, once we push the
expansion of the spectral density to one order higher in powers of $e^{-S_0}$,
i.e. the torus topology indicated in Eq.~\eqref{eq:JTspectralExpansion}. 

Referring to the original references~\cite{Eynard:2004mh, Saad2019} for details,
the expansion of the inverse Laplace transform of $Z(\beta)$ defines a
topological expansion of the resolvent $R(z)=\tr(G(z))$  of the
dual matrix theory as  
\begin{equation}
    R(z) = \sum_{g=0}^\infty  \frac{R_{g,1}(z)}{(e^{S_0})^{2g-1}}  \,,
\end{equation}
where $g$ labels the genera of surface geometries, and the subscript `1' in $R_{g,1}$ indicates that we are considering a single boundary.

 It has been shown~\cite{Eynard:2004mh,Saad2019,Eynard:2007fi} that the first terms of this expansion read
\begin{align*}
    R^{\rm JT}(z) = -e^{S_0}\frac{\sin(2\pi z)}{4\pi} - e^{-S_0} \frac{3 + 2\pi^2 z^2}{48 z^5} + {\cal O} \left(e^{-3S_0}  \right)\,.
\end{align*}
Taking the imaginary part, we obtain the density of states $\rho(E) = -\frac{1}{\pi}\textrm{Im} R(z)|_{z=\sqrt{E+i0}}$ as

\begin{eqnarray}
\label{eq:JT DoS 2nd order}
    \rho^{\rm JT}(E) &=& e^{S_0}\frac{1}{4\pi^2}\sinh\left(2\pi \sqrt{E} \right)  \\
    &&+ e^{-S_0}\left( \frac{1}{16\pi} E^{-5/2} + \frac{\pi}{24} E^{-3/2}\right) + {\cal O} \left( e^{-2S_0} \right). \nonumber
\end{eqnarray}
To relate this expression to our earlier discussion of the near edge spectral
density predicted by the flavor matrix model, we recall that the  asymptotic
scaling $\rho (\epsilon)= \frac{c}{\pi} \epsilon^{1 /2}$, defined the
dimensionless expansion parameter $\tilde \epsilon = \epsilon c^{2 /3}$, i.e.
energy in units of the effective level spacing. Comparison with Eq.~\eqref{eq:JT
DoS 2nd order} leads to $c=e^{S_0} /2$, and hence $\rho^\textrm{JT}(\epsilon)=
\frac{c^{2 /3}}{\pi}(\tilde \epsilon^{1/ 2}+ \frac{1}{32}\tilde \epsilon^{-5
/2})+\dots $, i.e. agreement with the expansion Eq.~\eqref{eq:DoSExpansion}. The
ellipses indicate  contributions proportional to $e^{- \alpha S_0}\tilde
\epsilon^\beta$, subleading in powers of the effective level spacing.  

To summarize, both the matrix model and JT gravity contain corrections to the
leading order spectral density $\sim \tilde \epsilon^{1 /2}$ which can be
captured by topological expansion. In either case, the next-to-leading order terms,
proportional to $\epsilon^{-5 /2}$ are described by terms of torus topology,
$g=1$. While in the gravitational path integral, these assume the form of a disk
 with a handle attached, cf. Eq.~\eqref{eq:JTspectralExpansion}, the
corresponding matrix ribbon diagram is shown in
Fig.~\ref{fig:SpectralDensityDiagram}. Either way, these diagrams represent the
leading order contributions to the expansion of the oscillatory Airy spectral
density Eq.~\eqref{eq:AiryKernel}. However,  such contributions are absent in the spectral
density of the SYK model, and likely any sparse model for that matter. 


The division between dense gravitational bulk systems and sparse quantum boundary theories
also shows up in probes of spectral correlations such as the form factor. While the
sparse form factor is subject to an average over collective fluctuations, cf.
Eq.~\eqref{eq:FormFactorRounding} and Fig.~\ref{fig:form}, the dense form factor
is not.

This section has highlighted physical differences between the behavior of dense
random matrix theories and sparsely random quantum chaotic systems, such as the
SYK model. These results show that it seems unlikely that there exist sparsely
encoded random systems whose chaotic signatures are compatible with those
predicted by bulk (pure) gravitational systems, such as JT gravity. Indeed,
the perturbative topological expansion of the latter is already known, \cite{Saad2019}, to be perturbatively equivalent to that of a dense
matrix model, and hence different to that of a  sparse quantum chaotic system. The
analysis here puts this observation on a more general footing: the 
spectra of sparse systems are subject to collective fluctuations
not present in dense systems, and it would thus appear
challenging to model pure low-dimensional gravity with a sparse 
boundary theory.

\section{Discussion}
In this paper we reasoned that the near-edge physics of  chaotic quantum systems
generically falls into one of two symmetry classes, sparse or dense. Which of
these is realized depends on the ratio between the number of a system's
independent random parameters and the Hilbert space dimension $D$,
logarithmic vs. algebraic  defining the two cases sparse vs. dense, respectively.
What both classes have in common is a non-analytic vanishing of the mean field
($D\to \infty$) spectral density $\bar \rho (\epsilon)\sim \epsilon^\alpha$, in
line with the interpretation of the edge as a symmetry breaking quantum critical
point. However, they are described by different near-edge effective
`Ginzburg-Landau' theories namely the Kontsevich matrix model (dense), and
a nonlinear sigma model modulated by a scalar fluctuation field (sparse),
respectively. 

Dense systems maintain a quasi-crystalline degree of stiffness
throughout the spectrum, implying oscillatory fine structures modulating the
mean field spectral density. We discussed a number of different approaches
revealing these structures, namely supersymmetric matrix integrals,
diagrammatic perturbation theory, and gravitational path integrals. 

Conversely, in sparse systems a much smaller averaging parameter space  is
responsible for large sample-to-sample `collective' fluctuations in the spectral
density. Individual system representatives still feature level repulsion in a
sequence terminating in an extremal energy level. However, the position of these
terminal points is subject to large fluctuations (in $\ln D$ ), generating tails
in the average spectral density and eradicating oscillatory fine structures. We
considered the SYK model as a representative case study to demonstrate the
statistically independent presence of two channels of fluctuations, the above
collective fluctuations, and microscale fluctuations reflecting level repulsion.
As a precision test we considered the non-perturbative
structure of the spectral form factor and demonstrated parameter-free agreement
between exact diagonalization and the predictions of the two-fluctuation channel
effective theory. 

Finally, we reasoned that while typical few-body interacting many-particle
systems are sparse, low-dimensional gravity is  dense. This observation is
troubling inasmuch as it appears to limit the scope of the holographic
correspondence between gravitational bulks and quantum boundary theories
realized in terms of `natural' many body systems. This disparity must likely be
seen in the context of the still somewhat mysterious status of the ensemble
average in low-dimensional holography and calls further study. 

\textit{Acknowledgments}
A.A. and K.W.K are grateful for numerous discussions with M. Berkooz on the  interpretation of chord diagram statistics.  T.~M.~acknowledges financial support by Brazilian agencies CNPq and FAPERJ.
K.W.K is supported by the National Research Foundation of Korea (NRF) grant
funded by the Korea government(MSIT) (No.2020R1A5A1016518). A.A. and M.R. acknowledge support from the Deutsche Forschungsgemeinschaft (DFG) Project No. 277101999, CRC 183 (project A03). 
Antonio M. Garc\'ia-Garc\'ia is thanked for eigenvalue data for $N=28$, and
J. J. M. V. is
  supported in part by U.S. DOE Grant  No. DE-FAG-88FR40388. 
\textbf{Data and materials availability:} Raw data, data-analysis, and codes used in the generation of Fig.5,7,8 are available in Zenodo with the identifier 10.5281/zenodo.10827097~\cite{zenodo}.

\bibliography{EdgeLibrary}

\begin{thebibliography}{47}
\bibitem [{foo()}]{footnote1}%
  \BibitemOpen
  \href@noop {} {}\bibinfo {note} {That is, the SYK Hamiltonian with $N=18$
  Majorana fermions acting on $D=2^8$-dimensional Hilbert space of fixed
  parity.}\BibitemShut {Stop}%
\end{thebibliography}

\appendix

\section{Matrix theory details}
\label{sec:MatrixTheoryDetails}

For the sake of completeness, we here review the construction of the `flavor
dual' of a Gaussian distributed `color' random matrix model (for a more detailed
discussion, see Ref.~\cite{Efetbook}).  We consider an ensemble of Gaussian distributed random Hamiltonians,
$H=\{H_{\mu \nu}\}$, $\mu,\nu=1 , \dots,D$ with
second moment
\begin{align}
    \label{eq:SecondMoment} 
    \left \langle H_{\mu \nu}H_{\nu' \mu'} \right \rangle\equiv \frac{\lambda^2}{D} 
    \delta_{\mu \mu'}\delta_{\nu \nu'}.
\end{align}
Our starting point is the Gaussian integral
\begin{align}
    \label{eq:GaussianFunctional}
    &Z(\hat \epsilon)=\int d \psi\, \left\langle\exp(-S[H,\psi,])\right\rangle,\cr 
    &\quad S[H,\psi]\equiv  i \bar \psi (\hat \epsilon-H) \psi,
\end{align}
where 
$\psi=\{\psi_\mu^\alpha\}$, $\alpha=(a,s)$ are complex commuting ($\psi^{\textrm{b},s}$) or Grassmann valued ( $\psi^{\textrm{f},s}$ ) integration variables. In this way it is guaranteed that for $\epsilon^{\textrm{b},s}=\epsilon^{\textrm{f},s}$ the integral is unit-normalized. 
  The conjugate variable $\bar \psi\equiv \psi^\dagger
\tau_3$ contains a Pauli matrix in the space of advanced and retarded indices
($\mathrm{Im}\,\hat \epsilon = i \delta \tau_3$) safeguarding the convergence of
the integral. (For Grassmann variables, convergence is not an issue and $\psi^\dagger$ a symbol for a variable independent of $\psi$.)

Integrating over the $H$-ensemble, we obtain a quartic action
\begin{align}
    \label{eq:ActionQuartic}
    Z(\hat \epsilon)=\int d \psi\, \exp \left(i \bar \psi \hat \epsilon \psi - 
    \frac{\lambda^2}{2 D}\str(B^2) \right),
\end{align}
where the bilinear $B\equiv \tau_3^{1/
2}\psi_\mu \bar \psi_\mu \tau_3^{1 /2}$. We decouple
this term by a Hubbard-Stratonovich transformation, and do the Gaussian integral over $\psi$ to arrive at  
\begin{align}
    \label{eq:AAction}
    & Z(\hat \epsilon)=\int dA\,\exp(-S[A]),\cr 
    &\quad S[A]\equiv D\left(\frac{1}{2\lambda^2}  \str A^2 + \str\ln (\hat \epsilon-A)\right), 
\end{align}
where  $A=\{A^{\alpha \beta}\}$ is a four-dimensional graded matrix. 

The global factor $D$ upfront invites a stationary phase analysis. We first seek
a solution for identical energy arguments,   $\hat{\epsilon} = E + i \delta
\tau_3$, except for the important imaginary increment $\pm i \delta$. The
variational equation $\delta_A S[\bar A]=0$ then assumes the form
\begin{align*}
    \bar A =  \frac{\lambda^2}{\epsilon+i \delta \tau_3 - \bar A}\Rightarrow \bar A = \frac{E}{2}- i \tau_3 \lambda \left(1-\left(\frac{E}{2\lambda}\right)^2 \right)^{\frac{1}{2}}, 
\end{align*}
where the sign of $\delta$  determines the branch of the square root function.
For decreasing energy, $E$, the magnitude of the symmetry breaking imaginary
part diminishes, and eventually vanishes at the spectral edges $E=-2\lambda$. 
The edge  theory  is obtained by expansion of the action in fluctuations $A\to
\bar A + A$ around the lower edge configuration, $\bar A = -E /2=\lambda$ \cite{verbaarschot1984replica}.
Re-introducing a  matrix of general near-edge energy configurations,
$\hat{\epsilon} \to 2\lambda + \hat \epsilon$, an expansion to leading order in
fluctuations, then leads to $S[A]=-D\str\left( \frac{\hat \epsilon
A}{\lambda^2}+\frac{A^3}{3\lambda^3}\right)$, and a final rescaling $A\to A
\sqrt{\lambda}$ brings the action into the form of
Eq.~\eqref{eq:KontsevichAction}  with the coupling constant $c=D/ \lambda^{3/ 2}$.

\section{Collective spectral fluctuations from chord diagrams}
\label{sec:CordDiagrams}
\begin{figure}[htbp]
    \centering
    \includegraphics[width=0.6\columnwidth]{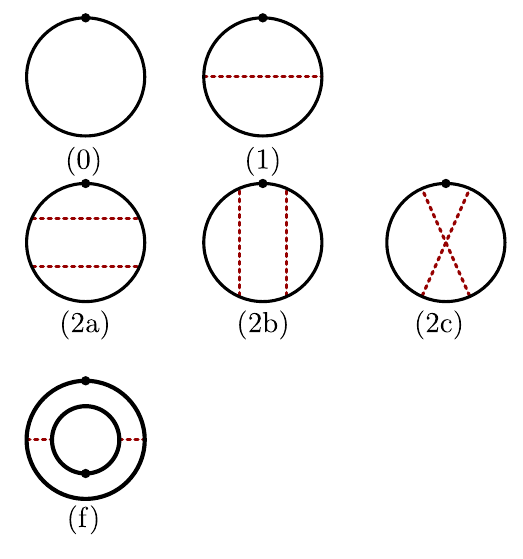}
    \caption{Chord diagrams of zeroth (0), second (1), and quartic order (2a-c) in $H$ contributing to the average spectral density. Diagram (f) describes the leading order contribution to the collective fluctuations, where the solid lines represent the summation of all diagrams contributing to the average spectral density.}
    \label{fig:CollectiveFluctuations}
\end{figure}

Referring to Ref.~\cite{Jia:2019orl}
detailed discussion, we here review how the average spectral density of the SYK
model, and its collective fluctuations are described in perturbation theory.
To start with, consider the configurational average of a single resolvent,
formally expanded in the Hamiltonian 
\begin{align}
\label{app_gf_expansion}
    \langle \tr\,G(z) \rangle=\sum_l z^{-2l+1}\langle \tr_\mathcal{F}(H^{2l}) \rangle.
\end{align}
We now average over the distribution \eqref{eq:JVariance} of the coupling
coefficients to generate a series, which up to fourth order 
reads as
\begin{align}
    \label{eq:CordExpansion}
    &\langle \tr\,G(z) \rangle= \frac{D}{z}+\frac{1}{z^3}\langle \tr_\mathcal{F}( H^2)\rangle +\frac{1}{z^5}\langle \tr_\mathcal{F}( H^4)\rangle+\dots \nonumber \\
&\quad  =\frac{D}{z}+\frac{\gamma^2}{z^3 K}\sum_\sigma\tr_{\mathcal{F}}(X_\sigma^2) \nonumber \\
&+\frac{\gamma^4}{z^5 K^2} \sum_{\sigma \sigma'} \tr_{\mathcal{F}}(X_\sigma^2 X_{\sigma '}^2+X_\sigma X_{\sigma '}^2 X_\sigma + X_\sigma X_{\sigma '} X_\sigma X_{\sigma '}) 
\nonumber \\
&\quad =\frac{D}{z}+\frac{D\gamma^2}{z^3}+\frac{D\gamma^4}{z^5} \left(2+\frac{1}{K^2}\sum_{\sigma,\sigma'}s_{\sigma,\sigma'}\right),
\end{align} 
with the sign factors defined through $X_\sigma X_{\sigma'}= s_{\sigma,\sigma'}X_{\sigma'}X_\sigma$, and where 
\begin{align}
    \label{eq:GammaDef}
    \gamma=\left(\frac{6K J^2}{N^3}\right)^{\frac{1}{2}}\approx \frac{N^{1 /2}J}{2}.
\end{align}
These few terms (cf. 
Fig.~\ref{fig:CollectiveFluctuations} for a visualization) already 
illustrate the principles of the perturbative analysis:
at $2^{\rm nd}$   order in the expansion, we have $(2n-1)!!$ contractions,
leading to different combinations, $\dots X_\sigma X_{\sigma'} X_{\sigma''}
\dots$, with pairwise occurrences of $ \dots X_\sigma \dots X_\sigma \dots\, $.
Reordering them to make use of the self-annihilation $X_\sigma^2=1$, we need
to permute operators $X_\sigma X_{\sigma'}$ (cf. diagram (2c), or
the last term in the third line of the above second order expansion) which
introduces sign factors $K^{-2}\sum_{\sigma \sigma'} s_{\sigma, \sigma'}=1-
\mathcal{O}(N^{-1})$, where the deviation of $\mathcal{O}(N^{-1})$ is due to the
few configurations where $\sigma$ and $\sigma'$ have an odd number of Majoranas
in common, and hence anti-commute.

Ignoring these corrections, the expansion assumes the form of an asymptotic
series, $\langle \tr\,G(z) \rangle \simeq  \frac{D}{z}\sum_{n} (\gamma
/z)^{2n}(2n-1)!!$. Its  combinatorial divergence can be dealt with
by Borel resummation \cite{verbaarschot1984replica}, i.e. we use the identity $(2n-1)!! =(2\pi)^{-1 /2}\int
dt\, e^{-t^2 /2}t^{2n}$, to represent the result as 
\begin{align*}
    &\langle \tr\,G(z) \rangle \approx \frac{D}{\sqrt{2\pi}z}\int dt\, e^{-\frac{t^2}{2}} \frac{1}{ 1-(\gamma t^2 /z )^2}
    \cr 
    &\qquad =-\frac{iD\sqrt{\pi}}{ \sqrt{2}\gamma} \exp\left(- \frac{z^2}{2 \gamma^2}
    \right),
\end{align*}   
where we assumed $\mathrm{Im}\, z>0$. With the identification $z=\epsilon^+$, we
conclude that to leading order in an $N^{-1}$-expansion, the spectral density is
Gaussian. The inclusion of corrections due to the above commutators
requires more work~\cite{garcia-garciaAnalyticalSpectralDensity2017}  and leads to the result \eqref{eq:AverageRhoSYK}. 

Finally, fluctuations of the DoS are described by the diagram (f), where the fat
lines represent the summation of all chords contributing to the average $\langle
\rho \rangle$. Relative to these, chords connecting between the two propagators
come at the expense of missing factors $K\gg1$, implying that it is sufficient
to work to the lowest non-vanishing order, two connecting chords. 


Following Ref.~\cite{Berkooz2021} the computation  of this diagram yields 
\begin{align}
\label{app_diagram_f}
{\rm var}(\rho(\epsilon))
&\approx
\frac{1}{2K}
[\partial_\epsilon
\epsilon\rho(\epsilon)]^2.
\end{align}
To see this, consider  the formal expansion Eq.~\eqref{app_gf_expansion} and
 note that non-vanishing diagrams $\langle {\rm tr}(H^k) {\rm tr}(H^m)
 \rangle_{\rm 2cc}$ have both of the connecting chords (2cc) locked to the same
 four-Majorana index $J_{\sigma}= J_{\sigma'}$. Otherwise, the leading order
 contribution to each individual product $H^k$ and $H^m$ contains the  unpaired
 Majorana operators $X_\sigma$ and  $X_{\sigma'}$ implying vanishing of their
 traces. Noting  that $H$ contains a summation over $K$ four-Majorana indices, this rationale leads to  \cite{Jia:2019orl}, $\langle {\rm tr}(H^k)
 {\rm tr}(H^m) \rangle_{\rm 2cc} \approx \frac{km}{2K} \langle{\rm
 tr}(H^k)\rangle \langle{\rm tr}(H^m)\rangle$, where the combinatorial factor
 accounts for the possibilities of choosing two connecting chords within each
 trace, $\frac{k}{2} \times \frac{m}{2}$, and the two ways of connecting the
 later. Using this approximation in the expansion Eq.~\eqref{app_gf_expansion},
 the estimate Eq.~\eqref{app_diagram_f} follows.

Linearizing on the other hand  
$\rho_\xi(\epsilon)=\frac{1}{1+\xi}\rho\left(\frac{\epsilon}{1+\xi}\right)$ 
in the spectral shift parameter, $\rho_\xi(\epsilon)\approx \rho(\epsilon)-\xi\partial_\epsilon \epsilon\rho(\epsilon)$, 
one finds   
\begin{align}
{\rm var}(\rho(\epsilon))
&\approx
\rm{var}(\xi)   
[\partial_\epsilon \epsilon\rho(\epsilon)]^2,
\end{align}
indicating that ${\rm var}(\xi)=1/(2K)$. Finally, neglecting contributions from derivatives of the DoS one arrives at 
Eq.~\eqref{eq:VarDoS} stated in the main text.

\section{Effective matrix theory of the SYK model}
\label{sec:EffectiveSYK}

\subsection{Derivation of the matrix action}

We here derive the effective matrix action
Eq.~\eqref{eq:ActionCollectiveFluctuations} describing the SYK model after
ensemble averaging \cite{altlandQuantumErgodicitySYK2018}. Our starting point is
again the Gaussian integral representation Eq.~\eqref{eq:GaussianFunctional},
only that the Hamiltonian is now the SYK Hamiltonian $H=\{H_{nm}\}$ in a first
quantized representation (i.e. interpreted as a matrix acting in a
$D$-dimensional Fock space, $\mathcal{F}$, in the fermion occupation number
basis $\{|n\rangle\}$) and the averaging is over the Gaussian distributed
coupling constants $J_a$  in Eq.~\eqref{eq:SYKHamiltonian}. Doing the averaging,
we obtain an action as in Eq.~\eqref{eq:ActionQuartic}, where the $B$-dependent
part now assumes the form 
\begin{align}
    \label{eq:BMatrixDef}
    \frac{3J^2}{ N^3}\sum_\sigma \STr(BX_\sigma B X_\sigma),\qquad B_{m n}=\tau_3^{\frac{1}{2}}\psi_m \bar \psi_n \tau_3^{\frac{1}{2}}. 
\end{align} 
Here, $(X_\sigma)_{m n}$ represent the Majorana quartets $X_\sigma$ in
Eq.~\eqref{eq:SYKHamiltonian} in terms of $D$-dimensional matrices, 
and the trace $\STr\equiv \, \tr_\mathcal{F}\,\str$
extends over both Fock-color and flavor space. We note
that the sparsity of the randomness shows in two complications: we
generate a sum over a large number of quartic terms indexed by the
$\mathcal{O}(N^4)$ labels $\sigma$, and the building blocks $B$  no longer are
`color singlets', but possess matrix structure in both color/Hilbert space
($m$-indices), and flavor space (the $\alpha$-indices implicit in $\psi$).

To mitigate the first of these problems, we use that~\cite{altlandQuantumErgodicitySYK2018} every  matrix
$O=\{O_{nm}\}$ in $D=2^{N /2}$-dimensional Fock space (we here consider the
entire Fock space of $N /2$ fermions, including both parity sectors) can be
expanded in the basis of $D^2$ matrices $X_\mu\equiv
\chi_{\mu_1}\chi_{\mu_2}\dots$, where the multi-index $\mu$ runs through all of
the $2^N=D^2$ possible combinations labeling distinct Majorana monomials,
$1,\chi_i,\chi_i \chi_j, \dots$. (We reserve the subscript $\sigma$ for 
matrices $X_\sigma$ containing the fixed number of $4$ operators, as in the SYK
Hamiltonian.) This expansion reads as~\cite{altlandQuantumErgodicitySYK2018}
\begin{align*}
    O=\frac{1}{D^{\frac{1}{2}}}\sum_\mu o_\mu X_\mu,\qquad o_\mu=\frac{1}{D^{\frac{1}{2}}}\tr_\mathcal{F}(O X_\mu^\dagger),
\end{align*}    
where we used the orthogonality
relation $\tr_\mathcal{F}(X_\mu X_\nu^\dagger)=\delta_{\mu \nu}D$, and the
coefficients $m_\mu =\{m_\mu^{\alpha \beta}\}$ are matrices in flavor space.

As a direct consequence of this representation change we obtain a decoupling
identity similar to a Fierz transform~\cite{garcia-garciaAnalyticalSpectralDensity2017},
\begin{align}
    \label{appendix_fierz_transformation}
    &\sum_\sigma{\rm tr}_{\cal F}\left(
    O X_\sigma O X_\sigma
    \right)
    =
    \frac{1}{D}\sum_\mu
    \sum_\sigma {\rm tr}_{\cal F}\left(
    X_\mu X_\sigma X_\mu X_\sigma
    \right)
    o_\mu^2.
    \end{align} 
Defining two sign factors through $s_\mu = X_\mu X_\mu$ and $X_\nu X_\mu =
    s_{\nu,\mu} X_\mu X_\nu$, we have ${\rm tr}_{\cal F}\left( X_\mu X_\sigma
    X_\mu X_\sigma \right)=D s_{\sigma,\mu}s_\mu$ and with 
\begin{align}
    \label{eq:SSignDef}
    S_\mu\equiv \frac{1}{K}\sum_\sigma s_{\sigma,\mu},
\end{align}
the trace identity assumes the form 
\begin{align*}
    &\sum_\sigma\STr\left(
    O X_\sigma O X_\sigma
    \right)
    =K
    \sum_\mu s_\mu S_\mu 
    o_\mu^2.
    \end{align*} 
We now apply this identity to the case where $O\to (3J^2 N^{-3})^{1 /2} B$
contains our bilinears in Eq.~\eqref{eq:BMatrixDef} with their internal flavor
matrix structure. As a result, we obtain
\begin{align*}
  \frac{3J^2  }{N^3}\STr(BX_\sigma B X_\sigma)
    = \frac{3J^2 K }{N^3}\sum_\mu s_\mu S_\mu \str(b_\mu^2)\equiv S[b].
\end{align*}
This representation is advantageous, in that it contains the essential
structure governing the expansion of the theory -- the sign factors $S_\mu$
describing the degree of non-commutativity of general Majorana monomials $X_\mu$
with the  quartets appearing in the Hamiltonian via Eq.~\eqref{eq:SSignDef} --
as Gaussian weights. However, it still contains the integration variables, in quartic order, via the bilinears $b_\mu$. 

To remedy   this latter problem, we perform a  Hubbard-Stratonovich decoupling in terms of  $D^2$ matrices
$a_\mu=\{a_\mu^{\alpha \beta}\}$. Using that $\str(a_\mu b_\mu)=\bar \psi
\tau_3^{\frac{1}{2}} a_\mu \tau_3^{\frac{1}{2}}\psi$, it is straightforward to
show that
\begin{align*}
    e^{-S[b]}=\int Da\, e^{-\frac{1}{2}\sum_\mu s_\mu S_\mu^{-1} \str(a_\mu^2)- i \gamma \bar \psi A \psi},
\end{align*}
where $A=\frac{1}{\sqrt{D}}\sum_\mu X_\mu a_\mu$,  $Da = \prod_\mu da_\mu$, the
coefficient $\gamma$ is defined in Eq.~\eqref{eq:GammaDef}, and we approximated
$K\approx N^4 /4!$. Finally, the integration over the now quadratic
$\psi$-dependence gets us to the effective action
\begin{align}
    \label{eq:EffectiveTrLnSYK}
    S[a]&=\frac{1}{2}\sum_\mu s_\mu S_\mu^{-1} \str(a_\mu^2)+\STr\ln \left(\hat \epsilon 
    +\gamma A \right),
\end{align}
where the trace of the logarithm extends over both color and flavor space. 

So far, all manipulations have been exact. As with the Gaussian distributed
matrix model, the action describing the ensemble average over SYK Hamiltonians
assumes the form of a quadratic term plus a tr ln. A stationary phase
approximation~\cite{altlandQuantumErgodicitySYK2018} would then lead to a
quadratic equation, and the prediction of a semicircular spectral density. While
this result is qualitatively correct in the band center, it fails at the band
edge. The problem with the stationary phase approach is that in the sparse
system, we have a large number of integration degrees of freedom in addition to the
single Fock-space homogeneous Goldstone mode $a_0$. While these may be individually
`massive', they do determine the profile of the spectral density, and in
particular describe the collective spectral fluctuations discussed in the main
text.

\subsection{Collective spectral fluctuations from the matrix action}

In the following, we show how to describe the joint effect of microscale correlations (the mode $a_0$) and collective fluctuations (all other modes $a_{\mu\not=0}$) in a unified manner starting from Eq.~\eqref{eq:EffectiveTrLnSYK}. We note that many of the concepts applied here to the SYK model  were pioneered in the paper~\cite{verbaarschot1984replica}.

For this purpose, it will be sufficient to start from a simplified version of the integral in
which we work only with the $s=+1$ sector. We consider
Eq.~\eqref{eq:CorrelationAExpectation} at $\alpha=1$, or
$\epsilon^1=\epsilon_1^+$ and   represent the derivative in $\epsilon_1$  at the
symmetric point $\epsilon_1^+=\epsilon_3^+\equiv \epsilon^+$, as
$\partial_\kappa|_{\kappa=0}$, where $\hat \epsilon \to \epsilon^+ \mathds{1}_2+
\kappa P$, and $P^{\alpha \beta}=\delta^{\alpha 1} \delta^{\beta 1}$ is a
projector onto the first of the four flavor indices. An expansion to the first order
in $P$ then is equivalent to taking the derivative.  

With this replacement, and $z=\epsilon^+$ we consider the expression
$G^+(z)=\partial_\kappa|_{\kappa=0}Z(\kappa)$, where $Z(\kappa)=\left \langle
\exp\left(- \STr\ln\left(z+ \kappa P +\gamma A  \right)\right) \right \rangle$,
and 
\begin{align*}
    \langle \dots \rangle =\int Da \, e^{-\frac{1}{2}\sum_\mu s_\mu S_\mu^{-1} \str(a_\mu^2)}(\dots).
\end{align*}
We aim to understand how the expansion in $\gamma$ relates to the chord diagram
expansion of the spectral density reviewed in section \ref{sec:CordDiagrams}. 
With $G_\kappa\equiv (z+ \kappa P)^{-1}$, we obtain
$Z=Z^{(0)}+Z^{(2)}+Z^{(4)}+\dots$, where
\begin{align*}
    Z^{(0)}(\kappa)&=\exp(-\STr\ln(G_\kappa)),\cr 
    Z^{(2)}(\kappa)&=\frac{\gamma^2}{2}\left \langle (\STr(G_\kappa A))^2+\STr((G_\kappa A)^2) \right \rangle,\cr 
    Z^{(4)}(\kappa)&=\frac{\gamma^4}{4} \left \langle \STr((G_\kappa A)^2) (\STr(G_\kappa A))^2+\STr((G_\kappa A)^4)  \right \rangle.
\end{align*} 
(The application of the Wick contraction outlined below to fourth-order
combinations such as $(\STr(G_\kappa A))^4$ or $\STr(G_\kappa A)\STr(G_\kappa A)^3$ leads to products of supertraces $\STr(\dots)\STr(\dots)\dots$, instead of a single $\STr$. After expansion to first order in $P$ we are left with factors $\STr(G_0)$. The proportionality of $G_0$ to unity in graded space implies the vanishing of these terms.)  We next need to
compute the integrals over the Gaussian weight, and to this end we use a Wick
contraction formula for the $2\times 2$-matrices $a_\mu$, which can be checked
by straightforward computation 
\begin{align*}
    \langle \str(a_\mu X a_\nu Y) \rangle= \delta_{\mu \nu}S_\mu s_\mu\, \str(X)\str(Y),\cr 
    \langle \str(a_\mu X) \str(a_\nu Y) \rangle=\delta_{\mu \nu}S_\mu s_\mu\, \str(XY). 
\end{align*}  
We also note that the proportionality $G_\kappa\propto \mathds{1}_\mathcal{F}$
implies $\STr(G_\kappa A)= D^{1 /2}\str(G_\kappa a_0)$. With these auxiliary
identities, we find $Z^{(0)}\to  -D z^{-1}\kappa$, where the arrow stands for
the first order expansion in $\kappa$, and we noted $\str(P)=1$. The first order
term becomes $Z^{(2)}=\frac{\gamma^2}{2}D \,\str(G_\kappa^2)\to -\frac{D
\gamma^2}{z^3}\kappa$, where we used $S_0=s_0=1$. Finally, the second order term
yields 
\begin{align*}
    Z^{(4)}&= \frac{\gamma^4}{4}\str(G_\kappa^4)\cr 
    &\qquad \left(2D+\sum_{\mu \nu }s_\mu s_\nu S_\mu S_\nu \frac{1}{D^2}\tr_\mathcal{F}(X_\mu X_\nu X_\mu X_\nu)\right)\to \cr 
    &\qquad \to - \frac{ \gamma^4 \kappa}{z^5}\left(2D+\frac{1}{D}\sum_{\mu \nu }s_\mu s_\nu S_\mu S_\nu s_{\mu,\nu}\right).
\end{align*}      
Using the definition \eqref{eq:SSignDef} one can verify that $\sum_{\mu \nu
}s_\mu s_\nu S_\mu S_\nu
s_{\mu,\nu}=\frac{D^2}{K^2}\sum_{\sigma,\sigma'}s_{\sigma,\sigma'}$. With these
results, we find that the expansion of the resolvent in $a$-mode fluctuations,
$\langle G(z) \rangle=\partial_\kappa Z(\kappa)$ is identical to that in chord
diagrams, Eq.~\eqref{eq:CordExpansion}.  More generally, it is straightforward
to verify that the Wick contraction $\langle \STr(X A Y A) \rangle$  of
$a$-insertions is equivalent to the insertion of an $H$-contraction between the
Fock space matrices $X$ and $Y$.      
 
The conclusion here is that individual $a$-contractions represent the chord
lines of the original SYK model. At the same time, it is not possible to compute
the $a$-integrals to arbitrary order and in closed form. Instead, we here adopt
a pragmatic approach. Consider the integration organized in such a way that we
integrate over fluctuations diagonal in causal space, \(a^{ss}\), first. In a
second step, we then consider \(a^{s \bar s}\), anticipating that generic
fluctuations coupling between different causal branches are suppressed in powers
of \(K\), cf. the reasoning at the end of Appendix \ref{sec:CordDiagrams}. The
exception to this rule are the modes \(a^{+-}_{0}\) and \(a^{-+}_{0}\),
isotropic in Hilbert space, which require a separate treatment. Turning back to
the modes \(a^{ss}\), it is straightforward, if tedious, to check that the
expansion of the tr ln, followed by integration against the Gaussian weight,
reproduces the chord diagram expansion. (As it should, so far, no approximations
have been made.) This observation has two consequences: first, after
resummation, the resolvent  Green function (i.e. the formal inverse of the
operator under the logarithm) is dressed by a `self energy' whose imaginary part
carries a sign factor \((-1)^s\). This is the formal statement of \emph{symmetry
breaking}. From a theory with infinitesimal \(i s \delta\), we proceed to one
with finite difference between the retarded and advanced contour. Second, the
imaginary part of that Green function yields the cord diagram spectral density,
discussed above.  


On this basis, we may then proceed to the integration over the modes
$a_0^{+-}$ and $a_0^{-+}$. Being isotropic in color space and off-diagonal in
$s$-space they are the Goldstone modes of the above symmetry breaking and must
be treated non-perturbatively.  To this end, we introduce an off-diagonal
$4\times 4$ matrix zero mode generator $W$, defined to anti-commute with the
symmetry breaking matrix, $[W,\tau_3]_+=0$. With  $T=\exp(W)$, we change
variables, $a_0 \to T(a_0^{++}\oplus a_0^{--})T^{-1}$. Likewise, $a_\mu \to T
a_\mu T^{-1}$, for all Hilbert space inhomogeneous modes, $\mu\not= 1$. In this
representation, the Goldstone mode generators enter in a form conceptually
similar to that of a
rotation  degree of freedom in a symmetry broken ferromagnetic phase (small  energy differences, \(\omega\)  playing the role of a symmetry breaking magnetic field, in this analogy).    
Furthering this analogy, we substitute
the new representation into \eqref{eq:EffectiveTrLnSYK}, and turn to a rotated
frame, 
\begin{align}
    S[a,T]&=\frac{1}{2}\sum_\mu s_\mu S_\mu^{-1} \str(a_\mu^2)+\STr\ln \left(T^{-1}\hat \epsilon T 
    +\gamma A \right),
\end{align}
where we noted the $T$-independence of the Gaussian weight, and the matrix $a_0$
now excludes the off-diagonal sector. Formally defining the effective Goldstone
mode action by an integral over the `radial degrees of freedom',
\begin{align*}
    \exp(-S[T])\equiv \langle \exp(-\STr\,\ln\left(T^{-1}\hat \epsilon T 
    +\gamma A \right)) \rangle,
\end{align*}    
where the angular brackets denote integration over the Gaussian weight, as
before, we now consider the limit $\hat \epsilon \equiv \epsilon
\mathds{1}+\hat{\omega}$, where the entries $|\omega^\alpha|\sim \rho^{-1}$ of
the explicitly symmetry breaking matrix $\hat \omega$ are of the order of the
average level spacing. With $T^{-1}\hat \epsilon T= \epsilon \mathds{1}+
T^{-1}\hat{\omega}T$, a first order expansion in $\hat{\omega}$ yields
$\exp(-S[T])\approx \left\langle \exp(-i\STr\left( \mathrm{Im}\,(G[A])
T^{-1}\hat{\omega} T \right))\right\rangle$, with $G[A]=(\epsilon + \gamma
A)^{-1}$. We now apply the above principle, and interpret
$\mathrm{Im}\tr_\mathcal{F}(G[A])=\pi \rho(\epsilon) \, \tau_3 $ as the
realization specific spectral density, and $\langle \dots \rangle$ as the
average over realizations. (As illustrated above, this identification can be
checked by explicit expansion in $a$-fluctuations.) 

Since the fluctuations $\mathrm{rms}(\rho(\epsilon))$  of the spectral density
are small compared to the average $\langle \rho(\epsilon) \rangle$ in the
parameter $K$, we may approximate 
\begin{align*}
    S[T]&=-i \pi \langle \rho(\epsilon) \rangle\, 
    \str(\tau_3 T^{-1}\hat\omega  T ) \\ 
    &\qquad
    +
    \frac{\pi^2\,\mathrm{var}(\rho(\epsilon))}{2}(\str(\tau_3 T^{-1}\hat\omega   T ))^2.
\end{align*}   
Substitution of $Q=T \tau_3 T^{-1}$, $\hat \omega= \frac{\omega}{2}\tau_3$,  and of Eq.~\eqref{eq:VarDoS} then leads to
Eq.~\eqref{eq:ActionCollectiveFluctuations}.

\section{Near edge spectral statistics}
\label{sec:DenseSpectralDetails}

In this Appendix, we provide details on the application of the Kontsevich model
Eq.~\eqref{eq:KontsevichAction} to the analysis of the near-edge spectrum. We
first consider the full model with four-dimensional $A$-matrices to compute
spectral correlations, and then reduce to a simpler two-dimensional variant to
obtain the average spectral density.  

\subsection{Spectral correlations}

We consider the action \eqref{eq:KontsevichAction} with the energy matrix
defined as $\hat{\epsilon}={\rm
diag}(\epsilon_1^+,\epsilon_2^-,\epsilon_3^+,\epsilon_4^-)=\epsilon+\frac{1}{2}\omega\tau_3+\alpha
\sigma_3 \tau_3$, where $\tau_3=\{(-1)^{s+1} \delta_{s s'}\}$  acts in graded
space, and $\sigma_3=\{(-1)^{a+1} \delta_{a a '}\}$ in causal space.   
From this representation, the full two-point correlation function, including
disconnected contributions, 
\begin{align}
    \label{eq:KCorrelationDef}
K(\epsilon_1 ,\epsilon_2 )
&=
\langle\rho(\epsilon_1 )\rangle
\langle\rho(\epsilon_2 )\rangle
+
C(\epsilon_1,\epsilon_2),
\end{align}
with $C(\epsilon_1,\epsilon_2)\equiv R_2(\epsilon_1,\epsilon_2) / \Delta^2(\epsilon)$, defined in Eq.~\eqref{eq:SpectralTwoPointDef} is obtained by differentiation, 
\begin{align}
\label{app:generating_function}
K(\epsilon_1 ,\epsilon_2 )
&\propto
\partial^2_\alpha 
\mathcal{Z}(\hat \epsilon)|_{\alpha=0},
\end{align}
from the partition sum Eq.~\eqref{eq:PartitionSum},
where we postpone the fixation of numerical prefactors to the final step of the computation.

Exploiting the symmetries of the action, we will compute the $A$-integral in a polar representation,
\begin{align}
A&=
iR P  R^{-1}, 
\quad 
P=
\begin{pmatrix}
P^+ & \\
& P^-
\end{pmatrix},
\quad 
P^s 
= 
\begin{pmatrix}
z^s 
& \\ 
& -w^s 
\end{pmatrix},
\end{align}
where the rotations $R$ contain all Grassmann variables, and the factor $i$ is
introduced such that the integration contours of the four radial coordinates $z^\pm, w^\pm$ are 
along the real axis, with an infinitesimal  shift into the upper complex half-plane for
convergence. In this parameterization, the generating
function becomes
\begin{align}
\label{app:generating_function_radial_coordinates}
Z(\hat \epsilon)
&=
\int dP J_P^2 
 e^{ic\,
{\rm str}(\frac{1}{3}P^3-\epsilon P) }
\int dR\, 
  e^{-ic\,{\rm str}(R^{-1} X R P)},
\end{align}
where $X=\frac{1}{2}\omega\tau_3 + \alpha \sigma_3 \tau_3$, $dP=\prod_s dz^sdw^s$, 
\begin{align}
J_P 
\propto 
\frac{(z^+-z^-)(w^+-w^-)}{(z^++w^+)(z^-+w^-)(z^++w^-)(z^-+w^+)}
\end{align}
is the supersymmetric generalization of the Vandermonde determinant,  
and $dR$ the Haar measure on the group of unitary supermatrices, $U(2|2)$.

The result of the $R$-integral is given by the Itzykson-Zuber  identity\cite{ItzyksonZuber94},
\begin{align}
\label{app:Itzykson-Zuber}
\int dR\, 
  e^{-ic\,
  {\rm str}(R^{-1} X R P)}
&=
\frac{\det(e^{-ic X^{{\rm b}s} z^{s'})} \det(e^{-ic X^{{\rm f}s} w^{s'}})}{J_X J_P}, 
\end{align}
with $X^{as}$, \(a=\textrm{b},\textrm{f}\), the diagonal elements of $X$ and
\begin{align*}
\qquad
J_X
\propto
\frac{(\omega+2\alpha)(\omega-2\alpha)}{
\alpha^2 \omega 
(\omega-2\alpha)}\to \alpha^{-2}, 
\end{align*}
where the arrow indicates that we need to retain only contributions
to $\mathcal{Z}(\alpha)$ of $\mathcal{O}(\alpha^2)$.   
Both, the determinant $J_P$, and the determinantal factors featuring in the
numerator of Eq.~\eqref{app:Itzykson-Zuber} are antisymmetric functions in the
variables $z^\pm$ and $w^\pm$, respectively. We may exploit this structure to replace the
latter as $\det(e^{-ic X^{{\rm b}s} z^{s'}}) \to e^{-ic \,\tr(
X^\textrm{b}P^\textrm{b})}$. (The anti-symmetrization implied by multiplication
with $J_P$ restores the determinant.) Proceeding in the same way with the fermionic determinant, the numerator of Eq.~\eqref{app:Itzykson-Zuber} is replaced by the factor $\exp(-ic \,\str(X P))\to \exp(- \frac{ic \omega}{2}\str(P \tau_3) )$, where the arrow indicates that we have already used up our two powers of $\alpha$ in the factor $J_X$ and hence may reduce $X\to \omega \tau_3 /2$. Tidying up, and using Eq.~\eqref{app:generating_function}, we obtain
\begin{align*}
    K(\omega)
&\propto
\int dP J_P\, 
 e^{ic\,
{\rm str}\left( 
\frac{1}{3}P^3-\epsilon P 
-\frac{\omega}{2} P\tau_3
\right)}
\cr 
&\int dP\, 
J_P
e^{ic\left( 
S(z)+S(w)\right)},
\end{align*}
with 
\begin{align*}
S(x)
&=\sum_s \left(\frac{1}{3}x^{s3}
 - \left(\epsilon+ s \frac{\omega}{2}\right) x^s
\right).
\end{align*}
 To proceed, we note that 
\begin{align*}
J_P
&\propto
\frac{1}{
(z^++w^+)(z^-+w^-)
}
-
\frac{1}{
(z^++w^-)(z^-+w^+)
},
\end{align*}
express these fractions   as integrals,
\begin{align}
\label{app:fraction_integral_representation}
\frac{1}{iX}
&=
 \int_0^\infty dt\, 
 e^{-it X},
\end{align}
and recall the integral-representation of the Airy-function 
\begin{align}
\label{app:airy_integral_representation}
{\rm Ai}(x)
&= 
\int dz\, e^{i \left(\frac{1}{3}z^3 + x z \right)}.
\end{align}
With this we arrive at the final expression
\begin{align}
    \label{eq:KAiryResult}
K(\epsilon_1,\epsilon_2 )
&=
c^{\frac{4}{3}}
\left(
K_{\rm Ai}(x_1,x_1)
K_{\rm Ai}(x_2,x_2)
-
K_{\rm Ai}^2(x_1,x_2)
\right),
\end{align}
where $x_{1,2} \equiv -c^{\frac{2}{3}}\epsilon_{1,2}$, we reinstalled a normalization factor,  
\begin{align}
\label{eq:AiryKernel}
    K_{\rm Ai}(x,y)
\equiv \frac{{\rm Ai}(x){\rm Ai}'(y)-{\rm Ai}(y){\rm Ai}'(x)}{x-y},
\end{align}
is the Airy kernel, and we used the integral representation $K_{\rm Ai}(x,y)=\int_0^\infty dt\, {\rm Ai}(x+t){\rm Ai}(y+t)$. 


\subsection{Average spectral density}

The average spectral density can be calculated from a reduced Kontsevich model
in terms of two-dimensional (super)matrices~\cite{AltlandSonner21}. Choosing $\hat{\epsilon}=\epsilon +
\alpha \sigma_3$, we use that  $\rho(\epsilon)
\propto
\partial_\alpha 
Z(\hat \epsilon)|_{\alpha=0}$.
Here, $Z(\hat \epsilon)$ is given by Eq.~\eqref{eq:PartitionSum}, where the integration is over two-dimensional matrices lacking a causal structure. We  parameterize these as
\begin{align}
A&=
iRPR^{-1},
\qquad
P=
\begin{pmatrix}
z & \\ & -w
\end{pmatrix},
\end{align}  
to obtain  Eq.~\eqref{app:generating_function_radial_coordinates}, where now
$X=\alpha \sigma_3$, $dP=dz dw$, $dR$ is the Haar measure of $U(1|1)$, and $J_P
\propto (z+w)^{-1}$.
 Integration over the unitary supergroup the  
 Itzykson-Zuber integral identity with $J_X\propto 1/\alpha$ gives
 \begin{align}
 \rho(\epsilon)
 &\propto
 \int dP\, J_P e^{ic \left( S(z)+S(w)\right)},
 \end{align}
 where $S(x)
 =
 \frac{1}{3}x^3-\epsilon x$. We again represent $J_P$  an integral,
Eq.~\eqref{app:fraction_integral_representation}, recall the integral
representation  Eq.~\eqref{app:airy_integral_representation} of the Airy
function, and use that 
\begin{align}
K_\textrm{Ai}(x,x)=-x {\rm Ai}^2(x) + \left({\rm Ai}'(x)\right)^2.
\end{align} 
(This identity follows from the fact that $\textrm{Ai}$ solves the Airy
differential equation, $y^{\prime\prime}-x y=0$ and Taylor expansion in
Eq.\eqref{eq:AiryKernel}.) Upon restoring normalization, this leads to
$\rho(\epsilon)=c^{\frac{2}{3}} K_\textrm{Ai}(x)$  which is
Eq.~\eqref{eq:SpectralDensityEdgeDense}. We may also use this result to remove
the disconnected terms in the definition Eq.~\eqref{eq:KCorrelationDef}, i.e.
the first two terms in the result \eqref{eq:KAiryResult}, and arrive at Eq.~\eqref{eq:CorrelationFunctionAiry} for the connected correlation function.

\section{Kontsevich matrix model}

In this appendix, we apply diagrammatic perturbation theory to compute the
spectral density of the Kontsevich matrix model defined by
Eq.~\eqref{eq:PartitionSum} and Eq.~\eqref{eq:KontsevichAction} with $c=e^{s_0}/
2$. The  parameter controlling this expansion is $\tilde \epsilon^{1 /2} =
\epsilon^{1 /2} c^{1 /3}\gg 1$, cf. Eq.~\eqref{eq:SpectralDensityEdgeDense}, and
we will push the expansion to next to leading order beyond the mean field result
$ \langle \rho(\epsilon) \rangle\sim  \tilde \epsilon  \sim
\epsilon^{1 /2} $. As stated in the main text, the purpose of this
exercise is to develop some intuition for the matrix-theory scattering processes
responsible for generating structure in the spectral density.

\begin{figure}[h]
    \centering
    \includegraphics[scale=.9]{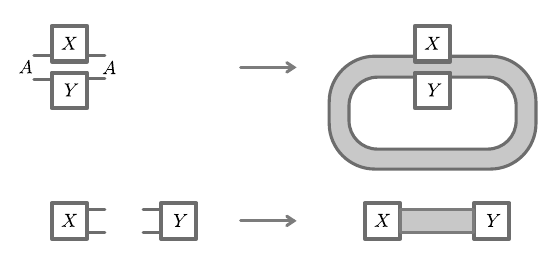}
    \caption{Diagrammatic representation of the Wick contraction rules of the
    $A$-matrix theory. Physically, the gray shaded areas correspond to
    ladder series of sequential scatterings off the random
    Hamiltonian~\cite{Altland2023}. }
    \label{fig:ContractionRules}
\end{figure}

\begin{figure}[h]
    \centering
    \includegraphics[scale=.9]{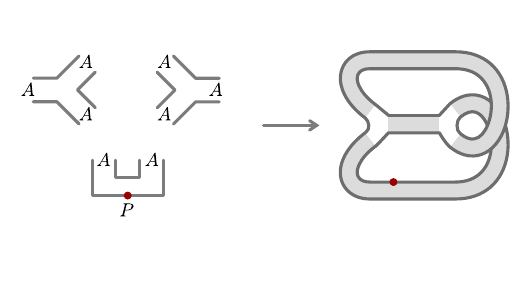}
    \caption{The leading fluctuation diagram modifying the mean field spectral
    density. Note the torus topology of this diagram: while its
    representation on a sheet of paper necessarily contains line crossings, a
    non-crossing representation on a torus is possible. Also note the
    resemblance to the diagrams of  appearing in, e.g., the periodic orbit
    theory of quantum chaos~\cite{heuslerPeriodicorbitTheoryLevel2007}. However,
    while these describe the co-propagation of amplitudes of different
    causality, we here have single self-retracing loops whose existence is tied
    to the spectral edge.  }
    \label{fig:SpectralDensityDiagram}
\end{figure}

The spectral density is calculated via Eq.~\eqref{eq:RhoResolvent} and Eq.~\eqref{eq:CorrelationAExpectation} from a functional integral with $2\times 2$-matrices (No retarded-advanced structure is required for the computation of the average spectral density)  as 
\begin{align*}
    \rho(\epsilon)= \frac{1}{\pi}\textrm{Im}\, \partial_{\epsilon_\textrm{b}}\big|_{\epsilon_\textrm{b}=\epsilon_\textrm{f}}\mathcal{Z}(\hat \epsilon),
\end{align*}
where $\hat{\epsilon}=\textrm{diag}(\epsilon_\textrm{b},\epsilon_\textrm{f})$. 
From the stationary phase analysis in section \ref{subsec:Stationary
 phase} for this action we obtain
\begin{align}
    \begin{split}
       \mathcal{Z} (\hat \epsilon) = e^{\frac{2i  c}{3}  
       \str(\hat{\epsilon}^{3 /2})}
        \int dA  \exp\left(c\, \str\left(i\hat{\epsilon}
        ^{1 /2} A^2 +\frac{1}{3}A^3\right)\right).
    \end{split}
\end{align}
The matrix structure of $A$ is defined in Eq.~\eqref{eq:GradedStructure}.
For $\epsilon_\textrm{b}=\epsilon_\textrm{f}$, $\str(\hat \epsilon^{3/ 2})=1$,
and the integral over $A$ is unit-normalized due to
supersymmetry~\cite{Efetbook}. With
$\partial_{\epsilon_\textrm{b}}|_{\epsilon_\textrm{b}=\epsilon_\textrm{f}}\hat
\epsilon^\alpha = \alpha  \epsilon^{\alpha-1}P_\textrm{b}$ and
$P_\textrm{b}=\textrm{diag}(1,0)$ a projector onto the boson subspace, the
differentiation of the integral then yields 
\begin{align*}
    \rho(\epsilon)=\frac{c^{2 /3}}{\pi}\left( \tilde \epsilon^{1 /2}+\frac{1}{2\tilde \epsilon^{1 /2}}\textrm{Re} \langle \str(A^2 P_\textrm{b}) \rangle  \right),
\end{align*}   
where
\begin{align*}
    \langle \dots \rangle=\int dA \, \exp\left( \,\str(i \tilde\epsilon^{1 /2}A^2 + A^3)  \right)(\dots),
\end{align*} 
and we rescaled, $A \to A c^{-1/ 3}$, to isolate our expansion parameter $\tilde
\epsilon$. To compute the fluctuation contribution, we define the Gaussian
average, $ \langle \dots \rangle_0=\int dA \, \exp(i \tilde\epsilon^{1 /2}\str\,
A^2)(\dots)$, and note the following Wick contraction rules (see Fig.~\ref{fig:ContractionRules})
\begin{align*}
    \langle \str(A X A Y) \rangle_0&= \frac{i}{ 2\tilde \epsilon^{1 /2}}\str(X)\,\str(Y),\cr 
    \langle \str(A X)\,\str( A Y) \rangle_0&= \frac{i}{ 2\tilde \epsilon^{1 /2}}\str(XY). 
\end{align*}  
With these identities, we obtain $\langle \str(P_\textrm{b}A A) \rangle_0
\propto \str(P_\textrm{b})\str(\mathds{1})=0$, and so we need to expand the
action in its nonlinearity. The first non-vanishing contribution is given by 
\begin{align*}
   & \langle \str(A^2P_\textrm{b})  \rangle\approx \frac{1}{2\cdot  3^2} \langle \str(A^2P_\textrm{b}) (\str(A^3))^2\rangle_0 = \cr
   &\quad=\frac{1}{16 \tilde \epsilon^{2}},
\end{align*}
where the above diagrammatic code implies the 'orbit representation' shown in
Fig.~\ref{fig:SpectralDensityDiagram}. Substituting this result into the formula
for the spectral density, we obtain
\begin{align*}
    \rho(\tilde \epsilon)\approx \frac{c^{2 /3}}{\pi}\left( \tilde \epsilon^{1 /2}+\frac{1}{32\tilde \epsilon^{5 /2}}  \right)
\end{align*}
With the identification $\Delta_0=c^{2 /3}$ this agrees with the expansion
Eq.~\eqref{eq:DoSExpansion}. However, we now have a diagrammatic interpretation
of the leading order correction in terms of the two-loop diagram shown in
Fig.~\ref{fig:SpectralDensityDiagram}, which we will compare to the expansion of
the gravitational path integral in section \ref{sec:Gravity}.  

\end{document}